# Impact of the WHO's 90-70-90 Strategy on HPV-Related Cervical Cancer Control: A Mathematical Model Evaluation in China

Hua Liu[a], Chunya Liu[a], Yumei Wei[b,*], Qibin Zhang[c], Jingyan Ma[d]

[a] School of Mathematics and Computer Science, Northwest Minzu University, Lanzhou, Gansu 730030, China
[b] Experimental Teaching Department, Northwest Minzu University, Lanzhou, Gansu 730000, China
[c] Gansu High-Tech Innovation Service Center, Lanzhou, Gansu 730000, China
[d] School of Preparatory Education Northwest Minzu University, Lanzhou, Gansu 730000, China

## Abstract

In August 2020, the World Health Assembly approved the Global Strategy to eliminate cervical cancer, marking the first time that numerous countries committed to eliminating a form of cancer. China introduced the HPV vaccine in 2016 and has made significant advancements in both prevention and treatment strategies. However, due to the relatively late introduction of the vaccine, the burden of cervical cancer in China continues to rise. In light of this, we develop a compartmental model to assess the impact of the WHO's 90-70-90 strategy, along with adult catch-up vaccination, on the control of HPV-induced cervical cancer in China. We analyze the basic properties of the model and provide proofs of the local and global asymptotic stability of the equilibrium points. Additionally, a sensitivity analysis is performed, and we use the MCMC algorithm to fit the number of new cervical cancer cases and deaths in China from 1990 to 2021. The estimated basic reproduction number before and after the introduction of the HPV vaccine in China is 1.5026 (95% CI: 1.4051-1.6002) and 1.0726 (95% CI: 0.9384-1.2067), respectively. The sensitivity analysis reveals that screening, as a non-pharmaceutical intervention, plays a crucial role in controlling the spread of the disease. We apply the 90-70-90 strategy to predict the future number of new cervical cancer cases and deaths in China. The results indicate that prioritizing the 70-90 target combination is the most cost-effective approach and can achieve the goal of zero new cervical cancer cases by 2061. Finally, an optimal control model is developed to explore the best implementation strategies for HPV vaccination and screening under various plausible scenarios.

## 1 Introduction

Cervical cancer is the fourth most prevalent cancer in women globally, mainly resulting from persistent infection with high-risk human papillomavirus (HPV) [1]. Although the early symptoms of cervical cancer are often subtle, as the disease progresses, it can lead to severe health consequences, including infertility, pain, and even death [2]. Globally, and particularly in developing countries and regions, cervical cancer incidence and mortality rates remain elevated due to limited healthcare resources and inadequate screening coverage [3], posing a significant public health threat. In 2020, more than 600,000 women globally were diagnosed with invasive cervical cancer, and over 300,000 women died from the disease [4]. China accounts for approximately one-fifth of the global cervical cancer burden, with an estimated 109,741 new cases reported in that year alone [5].

---

* Corresponding author.
  E-mail address: 7783360@qq.com, jslh@xbmu.edu.cn

The primary methods for preventing cervical cancer are HPV vaccination and regular cervical cancer screening. The HPV vaccine is most effective when administered before an individual is exposed to human papillomavirus, typically before the onset of sexual activity [6-7]. Therefore, the World Health Organization (WHO) recommends vaccinating girls aged 9 to 14. Vaccination at this age induces a stronger immune response and ensures protection before any potential exposure to the virus [8].

Cervical cancer screening coverage in China remains a significant challenge, with estimates indicating that only about 36.8% of women aged 35 to 64 have been screened at least once in their lifetime, which is slightly below the global average [9]. Although the national screening program has improved awareness and participation, barriers still exist in reaching higher screening rates. Compounding this issue, the HPV vaccine, which was first introduced in several countries in 2006, was not made available in China until 2016 [10]. This nearly decade-long delay meant that approximately 114 million girls aged 9 to 14 missed the optimal vaccination window, which is critical for ensuring protection before any potential exposure to HPV through sexual activity [11]. Consequently, these girls are at a heightened risk of HPV infection, increasing their likelihood of developing cervical cancer and related diseases in the future.

In May 2018, WHO Director-General Dr. Tedros, launched a global call to eliminate cervical cancer, highlighting that HPV vaccination for prevention and early treatment is a highly cost-effective strategy [12]. In August 2020, the World Health Assembly adopted the *Global Strategy to Accelerate the Elimination of Cervical Cancer*, with the goal of achieving, by 2030, a 90% vaccination rate among girls by age 15, 70% of women screened with high-performance tests by ages 35 and 45, and 90% of women diagnosed with cervical disease receiving appropriate treatment (90-70-90 strategy) [13]. By the end of 2023, 143 member states had introduced HPV vaccination into their national immunization programs, marking significant progress in cervical cancer prevention [14]. However, achieving this global target will require substantial efforts in policy-making, resource allocation, and public health education across countries.

Over the past few years, China has actively implemented global strategies to enhance cervical cancer prevention and control. In December 2020, the Chinese government of China announced its support for the *Global Strategy to Accelerate the Elimination of Cervical Cancer* [15]. This commitment involves advancing the global goal of eliminating cervical cancer through a clear three-tiered approach of vaccination, screening, and treatment. In January 2023, in line with the *Healthy China 2030 Plan* and the *Outline of Women's Development in China (2021–2030)*, ten departments, including the National Health Commission, the National Disease Control Bureau, and the National Medical Products Administration, jointly released the *Action Plan for Accelerating the Elimination of Cervical Cancer (2023-2030)* [16]. This plan set concrete targets, including achieving a cervical cancer screening rate of 50% among women of eligible age by 2025, and a treatment rate of 90% for patients with cervical cancer and precancerous lesions. The plan emphasizes efforts over the next seven years to expand HPV vaccination coverage, increase cervical cancer screening rates, aiming to significantly reduce the incidence and mortality of cervical cancer nationwide. Provinces have actively responded to this plan as well. For instance, Gansu Province's Health Commission issued the *Gansu Provincial Action Plan for Accelerating the Elimination of Cervical Cancer (2023-2030)* to help accelerate the elimination of cervical cancer and improve women's health across the province [17].

Since the WHO Director-General launched the global initiative to eliminate cervical cancer, governments, research institutions, policymakers, and public health experts have taken concrete actions to advance this goal [18]. Throughout this process, numerous mathematical models have been widely applied to the study of HPV infection and cervical cancer, particularly in the development of vaccination strategies and screening programs. Choi et al. [19] developed mathematical models to evaluate the cost-effectiveness of different cervical cancer screening strategies for both vaccinated and unvaccinated women in Hong Kong. The findings suggest that, for vaccinated cohorts, extending the screening intervals or reducing the number of screenings may be more cost-effective. Liu et al. [20] developed a mathematical model to study HPV transmission, incorporating the influence of media on public behavior. The model explores how media coverage, combined with vaccination and prevention measures, can reduce infection rates. Through stability and sensitivity analyses, the study demonstrates that enhancing media coverage can significantly aid in controlling HPV spread. Goldie et al. [21] developed a model to assess the cost-effectiveness of implementing HPV vaccination and cervical cancer screening across different regions. The study demonstrated that combining HPV vaccination with routine screening can significantly reduce both the incidence and mortality of cervical cancer. These findings provide important insights for shaping public health strategies on a global scale. Brisson et al. [22] proposed a strategy based on a mathematical model to evaluate the impact of HPV vaccination programs across different age groups. The results indicated that widespread HPV vaccination among girls and young women could significantly reduce the long-term incidence of cervical cancer. Hailegebireal et al. [23] analyzed the current state and influencing factors of cervical cancer screening in sub-Saharan African countries. The study revealed that screening coverage in the region falls significantly short of the 90-70-90 strategy targets, with multiple factors such as education level, economic status, and geographic location contributing to the low rates. We do not repeat previous research findings here; readers interested in those can refer to the relevant

studies[24-25].

Despite significant progress in HPV and cervical cancer modeling studies, several pressing issues remain unresolved. HPV vaccination programs are largely dependent on country-specific factors, with the key determinants being economic and geographical constraints, as well as the organization of healthcare systems [26]. Current modeling research primarily focuses on optimizing vaccination and screening strategies, but there is limited evaluation of adult HPV vaccine catch-up programs and long-term prevention outcomes. Moreover, how to effectively integrate mathematical modeling research with the WHO's 90-70-90 strategy is a topic that warrants further exploration.

Here, we focus on the long-term prediction and control of cervical cancer in China under the 90-70-90 strategy, evaluating how to further reduce or even eliminate new cases after the targets of the 90-70-90 strategy are met. By integrating mathematical models with health strategies, this study aims to reveal how HPV vaccination and screening programs can be effectively implemented and optimized in China's unique socio-economic context, thereby identifying the most optimal approach. In addition, we conducted a stability analysis of the model's equilibrium points and performed a sensitivity analysis on the model.

## 2 The HPV-induced Cervical Cancer Model

HPV, as a sexually transmitted infection, primarily spreads through intimate skin-to-skin contact. In most cases, the immune system clears the virus naturally within two years [18]; however, in some individuals, especially those with weakened immune systems or certain risk factors, HPV can persist. Cervical cancer is primarily caused by persistent infection with high-risk HPV types, notably HPV 16 and 18. The process typically begins with the virus infecting the epithelial cells of the cervix. Over time, the viral DNA integrates into the host cell's genome, disrupting normal cellular functions. This leads to the development of precancerous lesions known as cervical intraepithelial neoplasia, which can progress from low-grade lesions to high-grade lesions if left untreated. If the precancerous changes are not detected and treated, they can evolve into invasive cervical cancer, characterized by the infiltration of cancerous cells into surrounding tissues [27]. This transformation can take years, often a decade or more, highlighting the importance of regular screening and early intervention to prevent the progression from HPV infection to cervical cancer.

Cervical cancer screening in China has seen significant developments over the past few decades, particularly following the launch of national initiatives aimed at improving women's health. The first major efforts began in the 1950s, but systematic screening programs were not widely implemented until the early 2000s[28]. In 2009, China initiated a large-scale cervical cancer screening program targeting rural populations, which has since expanded access to screening services across the country [29]. For individuals undergoing screening, including those who are susceptible to infection and those unaware of their HPV status, several potential outcomes may arise. A normal screening result indicates no abnormalities, which may include individuals who are unknowingly carrying HPV, as most HPV infections are transient;

approximately 70% resolve within a year, and 90% within two years [30]. Conversely, if abnormalities are detected, the results may categorize the findings as precancerous lesions, ranging from low-grade squamous intraepithelial lesions to high-grade lesions, indicating a higher risk of developing cervical cancer and necessitating further investigation. In some cases, screenings can lead directly to a diagnosis of cervical cancer, allowing for timely treatment.

Based on the descriptions above and the dynamics of HPV transmission, we develop a Kermack-McKendrick-type model that divides the total population $N(t)$ into vaccinated individuals $V(t)$, susceptible individuals $S(t)$, unaware HPV carriers $I_u(t)$, patients in the precancerous stage who have not yet developed cervical cancer $P(t)$, invasive cervical cancer patients $C(t)$, and recovered individuals $R(t)$. The total population consists of sexually active individuals aged 15 to 64 years, hence, $N(t) = V(t) + S(t) + I_u(t) + P(t) + C(t) + R(t)$.

The model assumes a constant rate, $\Lambda$, at which individuals enter the sexually active population. A portion of adolescent receives the HPV vaccine prior to entering the sexually active population, denoted as $u$. The HPV catch-up vaccination rate for adults is $\eta$, and the vaccine waning rate is represented by $\omega$. It is assumed that invasive cervical cancer does not transmit the disease; only unaware HPV carriers and patients in the precancerous stage can transmit the infection to susceptible individuals, with transmission rates defined as $\beta_1$ and $\beta_2$, respectively. The screening population includes susceptible individuals and unaware HPV carriers, with a screening rate denoted as $\theta$. Susceptible individuals who undergo screening return to the susceptible population. A proportion $\phi$ of unaware HPV carriers who are screened transition into the precancerous stage, while a proportion $\sigma$ develop into invasive cervical cancer patients. The proportion of individuals with normal screening results is $1-\phi-\sigma$. Patients in the precancerous stage recover at a rate of $\varepsilon$ and transition to invasive cervical cancer at a rate of $\tau$. The natural mortality rate is represented by $\mu$, and the mortality rate due to invasive cervical cancer is $d$. The flowchart representing this model is shown in Fig. 1, and the model is formulated through the following ordinary differential equations:

$$\begin{cases} \dfrac{dV}{dt} = u\Lambda + \eta S - (\omega + \mu)V, \\ \dfrac{dS}{dt} = (1-u)\Lambda + \omega V - \dfrac{\beta_1 I_u + \beta_2 P}{N}S - (\eta + \mu)S, \\ \dfrac{dI_u}{dt} = \dfrac{\beta_1 I_u + \beta_2 P}{N}S - (\theta + \mu)I_u, \\ \dfrac{dP}{dt} = \theta\phi I_u - (\varepsilon + \tau + \mu)P, \\ \dfrac{dC}{dt} = \theta\sigma I_u + \tau P - (d + \mu)C, \\ \dfrac{dR}{dt} = \theta(1-\phi-\sigma)I_u + \varepsilon P - \mu R. \end{cases} \quad (1)$$

The initial value condition for the model (1) are given as:
$$\{V(0), S(0), I_u(0), P(0), C(0), R(0)\} \subseteq [0, \infty). \quad (2)$$

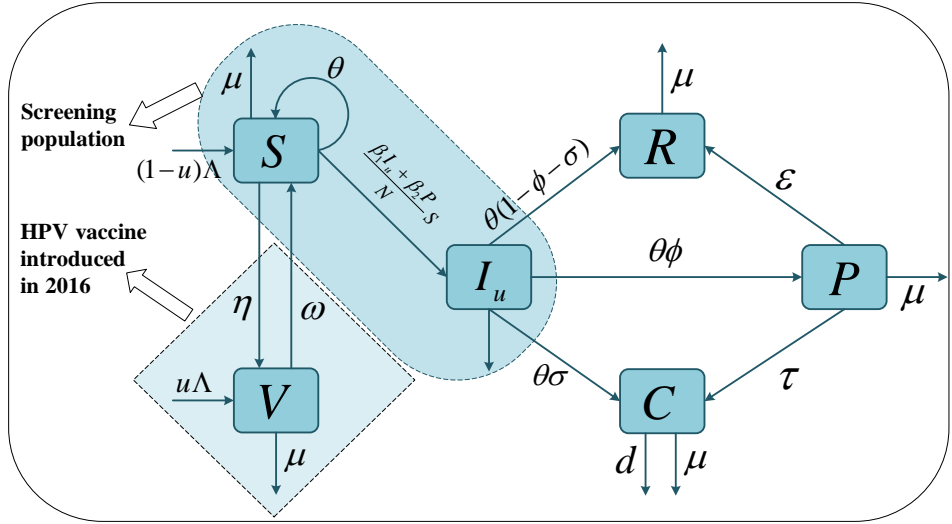

**Fig. 1** Schematic diagram of the HPV-induced cervical cancer model. The light blue boxes represent the vaccine compartments added after the introduction of the HPV vaccine in China in 2016, while the dark blue rectangular boxes represent the population undergoing cervical cancer screening.

Given that the HPV vaccine was not introduced in mainland China until 2016, to explore the dynamics of HPV prior to the vaccine's introduction, parameters $\eta, \omega, u$ and $V(t)$ in Model (1) can be set to 0. This paper aims to analyze the long-term prevention and control of cervical cancer under the 90-70-90 strategy in China, therefore, the theoretical analysis focuses on the dynamic behavior of Model (1).

**Theorem 1** The set
$$\Omega = \left\{ (V, S, I_u, P, C, R) \in \mathfrak{R}_+^6 : 0 \leq N = V + S + I_u + P + C + R \leq \frac{\Lambda}{\mu} \right\}$$

is positively invariant for Model (1).

**Proof** First, we prove that $V(t) \geq 0$, $S(t) \geq 0$, $I_u(t) \geq 0$, $P(t) \geq 0$, $C(t) \geq 0$ and $R(t) \geq 0$ with the initial condition (2) for all $t \geq 0$.

For the first equation of Model (1), we have
$$V(t) = V(0) e^{-\int_0^t (\omega+\mu) d\xi} + e^{-\int_0^t (\omega+\mu) d\xi} \int_0^t U(\xi) e^{\int_0^{\xi_1} (\omega+\mu) d\xi} d\xi_1, \quad \forall t \geq 0, \tag{3}$$

where $U(t) = u\Lambda + \eta S(t)$, therefore, $V(t) \geq 0$ when $V(0) \geq 0$ for all $t \geq 0$.

Let
$$W(t) = \min\{S(t), I_u(t), P(t), C(t), R(t)\}, \quad \forall t \geq 0. \tag{4}$$

Assuming that $W(t_1) = 0$, $t_1 \geq 0$. If $W(t) = S(t)$, so $I_u(t_1) \geq 0, P(t_1) \geq 0, C(t_1) \geq 0$ and $R(t_1) \geq 0$. It can be observed from the second equation of the Model (1) that:
$$\left. \frac{dS}{dt} \right|_{t=t_1} = (1-u)\Lambda + \omega V \geq 0.$$

Similarly, if $W(t)$ equals other state variables, then
$$\left. \frac{dI_u}{dt} \right|_{t=t_1} = \frac{\beta_2 P}{N} S \geq 0, \quad \left. \frac{dP}{dt} \right|_{t=t_1} = \theta \phi I_u \geq 0,$$

$$\left. \frac{dC}{dt} \right|_{t=t_1} = \theta \sigma I_u + \tau P \geq 0, \quad \left. \frac{dR}{dt} \right|_{t=t_1} = \theta(1-\phi-\sigma) I_u + \varepsilon P \geq 0.$$

Therefore, when the initial values $V(t) \geq 0$, $S(t) \geq 0$, $I_u(t) \geq 0$, $P(t) \geq 0$, $C(t) \geq 0$ and $R(t) \geq 0$, the solution of Model (1) belongs to $\mathfrak{R}_+^6$.

Next, we prove that the solution is bounded. By summing the equations of Model (1), we obtain:

$$\frac{dN}{dt} = \Lambda - \mu N - dC \leq \Lambda - \mu N, \tag{5}$$

according to standard comparison theorems, we get

$$N(t) \leq N(0)e^{-\mu t} - \frac{\Lambda}{\mu}e^{-\mu t} + \frac{\Lambda}{\mu}, \tag{6}$$

when $t \to \infty$, we have

$$N(t) \leq \frac{\Lambda}{d}.$$

Since $N(t) = V(t) + S(t) + I_u(t) + P(t) + C(t) + R(t)$, and based on the initial condition (2), we have $N(0) \geq 0$. Therefore, the region $\Omega$ is a positively invariant set for Model (1).

## 3 Model Analysis

### 3.1 Basic Reproduction Number

In this section, we use the next-generation matrix method to calculate the basic reproduction number $R_0$. We first calculate the disease-free equilibrium (DFE) point $E_0 = (V_0, S_0, 0, 0, 0, 0)$ for Model (1), where

$$V_0 = \frac{\Lambda(\mu u + \eta)}{\mu(\mu + \eta + \omega)}, \quad S_0 = \frac{\Lambda(\omega + \mu - \mu u)}{\mu(\mu + \eta + \omega)}.$$

The infected compartments in Model (1) are $I_u$, $P$ and $C$, arranged as $(I_u, P, C)$. The nonlinear expressions for new infections, represented by $\mathcal{F}$, and the outflow term, denoted as $\mathcal{V}$, are as follows:

$$\mathcal{F} = \begin{bmatrix} (\beta_1 I_u + \beta_2 P)S/N \\ 0 \\ 0 \end{bmatrix}, \quad \mathcal{V} = \begin{bmatrix} (\theta + \mu)I_u \\ (\varepsilon + \tau + \mu)P - \theta\phi I_u \\ (d + \mu)C - \theta\sigma I_u - \tau P \end{bmatrix}. \tag{7}$$

Taking the derivatives of $\mathcal{F}$ and $\mathcal{V}$, and substituting the DFE point, we have

$$F = \begin{bmatrix} \frac{\beta_1 S_0}{N} & \frac{\beta_2 S_0}{N} & 0 \\ 0 & 0 & 0 \\ 0 & 0 & 0 \end{bmatrix}, \quad V = \begin{bmatrix} \theta + \mu & 0 & 0 \\ -\theta\phi & \varepsilon + \tau + \mu & 0 \\ -\theta\sigma & -\tau & d + \mu \end{bmatrix}. \tag{8}$$

Therefore, $R_0$ is the spectral radius of the product of the $F$ and $V^{-1}$ matrices, that is,

$$R_0 = R_{I_u} + R_P = \frac{\beta_1(\omega + \mu - \mu u)}{(\theta + \mu)(\mu + \eta + \omega)} + \frac{\beta_2 \theta\phi(\omega + \mu - \mu u)}{(\theta + \mu)(\varepsilon + \tau + \mu)(\mu + \eta + \omega)}. \tag{9}$$

From the expression of $R_0$, it is clear that it consists of two components. The $R_{I_u}$ represents the number of secondary infections caused by individuals who are unaware they carry HPV and come into contact with susceptible individuals. The $R_P$ refers to the number of secondary infections resulting from contact between susceptible individuals and patients with precancerous lesions. Each term of $R_0$ has an epidemiological interpretation. The mean duration in $I_u$ is $1/(\theta+\mu)$ with contact rate $\dfrac{\beta_1(\omega+\mu-\mu u)}{(\mu+\eta+\omega)}$, giving a contribution to $R_0$ of $\dfrac{\beta_1(\omega+\mu-\mu u)}{(\theta+\mu)(\mu+\eta+\omega)}$. A fraction $\dfrac{\theta\phi}{(\theta+\mu)}$ goes from $I_u$ to $P$, with contact rate $\dfrac{\beta_2(\omega+\mu-\mu u)}{(\theta+\mu)(\mu+\eta+\omega)}$ and mean duration $\dfrac{1}{(\varepsilon+\tau+\mu)}$. The total of these individual contributions results in $R_0$.

Similarly, the same method can be used to calculate the basic reproduction number $R_0'$ for the model prior to the introduction of the HPV vaccine, namely,

$$R_0' = R_{I_u}' + R_P' = \frac{\beta_1}{(\theta+\mu)} + \frac{\beta_2\theta\phi}{(\theta+\mu)(\varepsilon+\tau+\mu)}. \tag{10}$$

**3.2 Stability of Disease-Free Equilibrium**

**Theorem 2** If $R_0 < 1$, the DFE $E_0$ of Model (1) is locally asymptotically stable.

**Proof** The Jacobian matrix of model (1) at the DFE point $E_0$ is a 6×6 matrix, where two eigenvalues, $-(d+\mu)$ and $-\mu$, can be easily identified. The remaining part of the matrix becomes a 4×4 matrix.

$$J_{E_0} = \begin{bmatrix} -(\omega+\mu) & \eta & 0 & 0 \\ \omega & -(\eta+\mu) & -\dfrac{\beta_1 S_0}{N} & -\dfrac{\beta_2 S_0}{N} \\ 0 & 0 & \dfrac{\beta_1 S_0}{N}-(\theta+\mu) & \dfrac{\beta_2 S_0}{N} \\ 0 & 0 & \theta\phi & -(\varepsilon+\tau+\mu) \end{bmatrix}. \tag{11}$$

The other four eigenvalues can be obtained as the roots of the following characteristic equation.

$$(\lambda+\mu)(\lambda+\eta+\mu+\omega)[(\lambda-\frac{\beta_1 S_0}{N}+(\theta+\mu))(\lambda+\varepsilon+\tau+\mu)-\frac{\beta_2 S_0}{N}\theta\phi]=0 \tag{12}$$

From Eq. (5), we can obtain two characteristic roots, $-\mu$ and $-(\eta+\mu+\omega)$. Let's continue to calculate the remaining characteristic equation, we have

$$\lambda^2 + b_1\lambda + b_2 = 0, \tag{13}$$

where $b_1 = (\varepsilon+\tau+\mu)-(\theta+\mu)(R_{I_u}-1)$, $b_2 = -(\theta+\mu)(\varepsilon+\tau+\mu)(R_0-1)$.

If $R_0 < 1$, we have $R_{I_u} < 1$, so $b_1 > 0, b_2 > 0$. According to the Routh–Hurwitz criterion, the roots of Eq. (6) are either negative or have negative real parts. Therefore, DFE point $E_0$ is locally asymptotically stable (LAS).

**Theorem 3** If $R_0 < 1$, the DFE $E_0$ of Model (1) is globally asymptotically stable.

**Proof** According to Huo et al. [31], we define the Lyapunov function as follows:

$$L(t) = \frac{(\omega+\mu-\mu u)}{(\mu+\eta+\omega)}I_u(t) + \frac{\beta_2(\omega+\mu-\mu u)}{(\varepsilon+\tau+\mu)(\mu+\eta+\omega)}P(t). \tag{14}$$

The time derivative of $L(t)$ is:

$$\frac{dL(t)}{dt} = \frac{(\omega+\mu-\mu u)}{(\mu+\eta+\omega)} \frac{dI_u(t)}{dt} + \frac{\beta_2(\omega+\mu-\mu u)}{(\varepsilon+\tau+\mu)(\mu+\eta+\omega)} \frac{dP(t)}{dt}$$

$$= \frac{(\omega+\mu-\mu u)}{(\mu+\eta+\omega)}[\frac{\beta_1 I_u + \beta_2 P}{N}S - (\theta+\mu)I_u]$$

$$+ \frac{\beta_2(\omega+\mu-\mu u)}{(\varepsilon+\tau+\mu)(\mu+\eta+\omega)}[\theta\phi I_u - (\varepsilon+\tau+\mu)P]$$

$$= \frac{(\omega+\mu-\mu u)}{(\mu+\eta+\omega)}[\frac{\beta_1 S}{N} - (\theta+\mu) + \frac{\beta_2\theta\phi}{(\varepsilon+\tau+\mu)}]I_u$$

$$+ \frac{(\omega+\mu-\mu u)}{(\mu+\eta+\omega)}[\frac{\beta_2 S}{N} - \frac{\beta_2}{(\varepsilon+\tau+\mu)}(\varepsilon+\tau+\mu)]P.$$

Since $S(t) \leq N(t)$, we have

$$\frac{dL(t)}{dt} \leq \frac{(\omega+\mu-\mu u)}{(\mu+\eta+\omega)}(\theta+\mu)[\frac{\beta_1}{(\theta+\mu)} + \frac{\beta_2\theta\phi}{(\theta+\mu)(\varepsilon+\tau+\mu)} - 1]I_u$$

$$= (\theta+\mu)[R_0 - \frac{(\omega+\mu-\mu u)}{(\mu+\eta+\omega)}]I_u$$

$$\leq (\theta+\mu)[R_0 - \frac{(\omega+\mu-\mu u)}{(\mu+\eta+\omega)} + \frac{\mu u+\eta}{(\mu+\eta+\omega)}]I_u$$

$$= (\theta+\mu)(R_0 - 1)I_u.$$

When $R_0 < 1$, we can conclude that $L'(t) < 0$. While $L'(t) = 0$, if and only if $(V(t), S(t), I_u(t), P(t), C(t), R(t)) = (V_0, S_0, 0, 0, 0, 0)$. According to LaSalle's invariance principle [32], the DFE $E_0$ of Model (1) is globally asymptotically stable (GAS).

We selected a set of parameters to plot the stability diagram of the DFE, thereby verifying the theoretical analysis above. The stability diagram of the DFE is presented in Supplementary Note 1.

### 3.3 Stability of Endemic Equilibrium

First, we determine the endemic equilibrium (EE) $E^* = (V^*, S^*, I_u^*, P^*, C^*, R^*)$, The detailed calculation process can be found in Supplementary Note 2, where

$$V^* = \frac{\Lambda(\mu u+\eta) - \eta(\theta+\mu)I_u^*}{\mu(\mu+\eta+\omega)}, \quad S^* = \frac{\Lambda(\omega+\mu-\mu u) - (\theta+\mu)(\omega+\mu)I_u^*}{\mu(\mu+\eta+\omega)}.$$

$$I_u^* = \frac{\Lambda(R_0 - 1)}{R_0 \frac{(\theta+\mu)(\omega+\mu)}{\omega+\mu-\mu u} - \frac{d\theta[\phi\tau + \sigma(\varepsilon+\tau+\mu)]}{(d+\mu)(\varepsilon+\tau+\mu)}},$$

$$P^* = \frac{\theta\phi}{\varepsilon+\tau+\mu}I_u^*, \quad C^* = \frac{1}{d+\mu}(\theta\sigma + \frac{\tau\theta\phi}{\varepsilon+\tau+\mu})I_u^*,$$

$$R^* = \frac{1}{\mu}[\theta(1-\phi-\sigma) + \frac{\varepsilon\theta\phi}{\varepsilon+\tau+\mu}]I_u^*.$$

With the equation for the variables $C(t)$ and $R(t)$ decoupled from Model (1), and since the decoupled model is mathematically equivalent to the original model, the following analysis ignore the equation for the variables $C(t)$ and $R(t)$.

**Theorem 4** When $R_0 > 1$, the EE $E^*$ of Model (1) exists and is LAS if the Routh–

Hurwitz criteria are satisfied.

**Proof** The Jacobian matrix of Model (1) is

$$J(E^*) = \begin{bmatrix} -(\omega+\mu) & \eta & 0 & 0 \\ \omega+a & a-b-(\eta+\mu) & -\dfrac{\beta_1 S}{N}+a & -\dfrac{\beta_2 S}{N}+a \\ -a & b-a & \dfrac{\beta_1 S}{N}-a-(\theta+\mu) & \dfrac{\beta_2 S}{N}-a \\ 0 & 0 & \theta\phi & -(\varepsilon+\tau+\mu) \end{bmatrix}_{|E^*}, \quad (15)$$

where

$$a = \frac{(\beta_1 I_u + \beta_2 P)}{N^2}S, \quad b = \frac{(\beta_1 I_u + \beta_2 P)}{N}. \quad (16)$$

Its characteristic equation is

$$\lambda^4 + B_1\lambda^3 + B_2\lambda^2 + B_3\lambda + B_4 = 0,$$

where

$$B_1 = \frac{1}{N}\Big[(\omega+\theta+\varepsilon+4\mu+\eta+\tau+b)N - \beta_1 S^*\Big],$$

$$B_2 = \frac{1}{N}\Big\{\big[6\mu^2 + 3(\omega+\theta+\varepsilon+\eta+\tau+b)\mu + (\omega+\varepsilon+\tau+\eta+(\phi-1)a+b)\theta$$
$$+(\varepsilon+b+\tau)\omega+(b+\eta)(\varepsilon+\tau)\big]N - \big((\eta+3\mu+\omega+\varepsilon+\tau)\beta_1 + \beta_2\theta\phi\big)S^*\Big\},$$

$$B_3 = \frac{1}{N}\Big\{\big[4\mu^3 + 3(\omega+\theta+\varepsilon+\eta+\tau+b)\mu^2$$
$$+2\big[(\omega+\varepsilon+\tau+\eta+(\phi-1)a+b)\theta+(\varepsilon+b+\tau)\omega+(b+\eta)(\varepsilon+\tau)\big]\mu$$
$$+\big((\varepsilon+\tau+b+(\phi-1)a)\omega+(-a+b+\eta)(\varepsilon+\tau)+a\eta(\phi-1)\big)\theta+\omega b(\varepsilon+\tau)\big]N$$
$$-\big((\eta+\mu+\omega)\mu\beta_1 + (\eta+2\mu+\omega)(\beta_1(\varepsilon+\tau+\mu)+\beta_2\theta\phi)\big)S^*\Big\},$$

$$B_4 = \frac{1}{N}\Big\{\big\{\mu^4 + (\omega+\theta+\varepsilon+\eta+\tau+b)\mu^3$$
$$+\big[(\omega+\varepsilon+\tau+\eta+(\phi-1)a+b)\theta+(\varepsilon+b+\tau)\omega+(b+\eta)(\varepsilon+\tau)\big]\mu^2$$
$$+\big[((\varepsilon+\tau+b+(\phi-1)a)\omega+(-a+b+\eta)(\varepsilon+\tau)+a\eta(\phi-1))\theta+\omega b(\varepsilon+\tau)\big]\mu$$
$$-\theta(\varepsilon+\tau)\big[(-b+a)\omega+a\eta\big]\big\}N - (\eta+\mu+\omega)\mu\big(\beta_1(\varepsilon+\tau+\mu)+\beta_2\theta\phi\big)S^*\Big\}.$$

Applying the Routh–Hurwitz criteria, if $B_1 > 0$, $B_1B_2 > B_3$, $B_1B_2B_3 > (B_3^2 + B_4B_1^2)$, and if $R_0 > 1$, then the EE $E^*$ of Model (1) is LAS.

Since $E_0$ lies on the boundary of set $\Omega$ and is unstable when the basic reproduction number $R_0 > 1$, this indicates that the Model (1) exhibits uniform persistence for $R_0 > 1$. In this case, certain state variables in Model (1) do not approach zero but remain bounded above a positive constant. Therefore, when $R_0 > 1$, there exists a constant $\zeta > 0$, independent of the initial conditions, such that with initial conditions $(V(0), S(0), I_u(0), P(0))$ in the interior of $\Omega$, the model's solution $(V(t), S(t), I_u(t), P(t))$ satisfies,

$$\liminf_{t\to\infty}\{V(t),S(t),I_u(t),P(t)\}\geq \zeta,\quad \text{for all } t. \tag{17}$$

The uniform persistence and boundedness of $\Omega$ imply the existence of a compact set $H$ within the interior of $\Omega$, which absorbs the dynamics of Model (1). Following this, we prove the GAS of $E^*$ using a geometric approach proposed by Li et al. [33]

**Theorem 5** If $H$ is a compact attracting set of the Model (1) on the region $\Omega$, and there exist a constant $\rho > 0$ and a Lozinski measure $\bar{\mu}$ such that $\bar{\mu}(Q) \leq -\rho$ for any $x \in H$, then every omega-limit sets of Model (1) in $\Omega$ serve as equilibrium points within $H$.

Based on the measure $\bar{\mu}$ outlined in Theorem 5, the estimation approach detailed in [34] is utilized. Define $\bar{\mu}(Q)$ as $\inf\{\psi : D_+ \|n\| \leq \psi \|n\|\}$, applicable to all solutions of $n' = Qn$. Here $D_+$ represents the derivative taken from the right-hand side. The objective is to establish a norm $\|\cdot\|$ within the $\mathfrak{R}^6$ space. This norm should ensure that for any element $x$ in $\Omega$, the measure $\bar{\mu}$ applied to the right-hand derivative of $x$, denoted as $D_+(x)$, consistently yields a negative value, i.e., $\bar{\mu}(D_+(x)) < 0$ for all $x \in \Omega$.

**Theorem 6** If $R_0 > 1$, the Routh–Hurwitz criteria are satisfied, and

$$\max\left\{\left(a(2+\frac{P}{I_u})-b(1+2\frac{P}{I_u})+\theta-(\omega+\eta+\mu)\right),\left(b-\mu-a\frac{P}{I_u}\right)\right\}<0,$$

then EE $E^*$ of Model (1) is GAS.

**Proof** The Jacobian matrix of Model (1) is

$$J = \begin{bmatrix} a_{11} & \eta & 0 & 0 \\ \omega+a & a-b+a_{22} & -\frac{\beta_1 S}{N}+a & -\frac{\beta_2 S}{N}+a \\ -a & b-a & \frac{\beta_1 S}{N}-a+a_{33} & \frac{\beta_2 S}{N}-a \\ 0 & 0 & \theta\phi & a_{44} \end{bmatrix}, \tag{18}$$

where

$$a_{11}=-(\omega+\mu),\ a_{22}=-(\eta+\mu),\ a_{33}=-(\theta+\mu),\ a_{44}=-(\varepsilon+\tau+\mu).$$

Then the second additive compound matrix of the Eq.(18) is

$$J^{[2]} = \begin{bmatrix} A_1 & -\frac{\beta_1 S}{N}+a & -\frac{\beta_2 S}{N}+a & 0 & 0 & 0 \\ b-a & A_2 & \frac{\beta_2 S}{N}+a & \eta & 0 & 0 \\ 0 & \theta\phi & A_3 & 0 & \eta & 0 \\ a & \omega+a & 0 & A_4 & \frac{\beta_2 S}{N}-a & \frac{\beta_2 S}{N}-a \\ 0 & 0 & \omega+a & \theta\phi & A_5 & -\frac{\beta_1 S}{N}+a \\ 0 & 0 & -a & 0 & b-a & A_6 \end{bmatrix}, \tag{19}$$

where

$$A_1 = a - b + a_{11} + a_{22}, \quad A_2 = \frac{\beta_1 S}{N} - a + a_{11} + a_{33}, \quad A_3 = a_{11} + a_{44},$$

$$A_4 = \frac{\beta_1 S}{N} - b + a_{22} + a_{33}, \quad A_5 = a - b + a_{22} + a_{44}, \quad A_6 = \frac{\beta_1 S}{N} - a + a_{33}.$$

We define a matrix function $M$ as follows,

$$M = \begin{bmatrix} N_1 & 0 & 0 & 0 & 0 & 0 \\ 0 & N_1 & 0 & 0 & 0 & 0 \\ 0 & 0 & 0 & N_1 & 0 & 0 \\ 0 & 0 & N_2 & 0 & 0 & 0 \\ 0 & 0 & 0 & 0 & N_2 & 0 \\ 0 & 0 & 0 & 0 & 0 & N_2 \end{bmatrix},$$

where

$$N_1 = \frac{1}{I_u}, \quad N_2 = \frac{1}{P}.$$

So,

$$M_f M^{-1} = -\text{diag}\{\frac{I_u'}{I_u}, \frac{I'}{I_u}, \frac{I'}{I_u}, \frac{P'}{P}, \frac{P'}{P}, \frac{P'}{P}\},$$

we can get,

$$M_f J^{[2]} M^{-1} = \begin{bmatrix} A_1 & -\frac{\beta_1 S}{N} + a & 0 & (-\frac{\beta_2 S}{N} + a)\frac{P}{I_u} & 0 & 0 \\ b - a & A_2 & \eta & (\frac{\beta_2 S}{N} - a)\frac{P}{I_u} & 0 & 0 \\ a & \omega + a & A_4 & 0 & (\frac{\beta_2 S}{N} - a)\frac{P}{I_u} & (\frac{\beta_2 S}{N} - a)\frac{P}{I_u} \\ 0 & \theta\phi\frac{I_u}{P} & 0 & A_2 & \eta & 0 \\ 0 & 0 & \theta\phi\frac{I_u}{P} & \omega + a & A_5 & -\frac{\beta_1 S}{N} + a \\ 0 & 0 & 0 & -a & b - a & A_6 \end{bmatrix}. \quad (20)$$

Let

$$Q = M_f M^{-1} + M_f J^{[2]} M^{-1}$$

$$= \begin{bmatrix} A_1 - \frac{I_u'}{I_u} & -\frac{\beta_1 S}{N} + a & 0 & (-\frac{\beta_2 S}{N} + a)\frac{P}{I_u} & 0 & 0 \\ b - a & A_2 - \frac{I_u'}{I_u} & \eta & (\frac{\beta_2 S}{N} - a)\frac{P}{I_u} & 0 & 0 \\ a & \omega + a & A_4 - \frac{I_u'}{I_u} & 0 & (\frac{\beta_2 S}{N} - a)\frac{P}{I_u} & (\frac{\beta_2 S}{N} - a)\frac{P}{I_u} \\ 0 & \theta\phi\frac{I_u}{P} & 0 & A_3 - \frac{P'}{P} & \eta & 0 \\ 0 & 0 & \theta\phi\frac{I_u}{P} & \omega + a & A_5 - \frac{P'}{P} & -\frac{\beta_1 S}{N} + a \\ 0 & 0 & 0 & -a & b - a & A_6 - \frac{P'}{P} \end{bmatrix}. \quad (21)$$

From Model (1), we have
$$\frac{I'_u}{I_u} = \frac{\beta_1 I_u + \beta_2 P}{N}\frac{S}{I_u} + a_{33}, \quad \frac{P'}{P} = \theta\phi\frac{I_u}{P} + a_{44}. \quad (22)$$

Referring to [35], the following norm is analyzed within the $\Re^6$ space,
$$\|n\| = max\{m_1, m_2\},$$
where $n = (n_1, n_2, n_3, n_4, n_5, n_6)^T$. We define two functions $m_1$ and $m_2$ to calculate the norm of $n$.

For $m_1(n_1, n_2, n_3)$, we have:
$$m_1(n_1,n_2,n_3) = \begin{cases} max\{|n_1|, |n_2|+|n_3|\}, & \text{if } sgn(n_1)=sgn(n_2)=sgn(n_3), \\ max\{|n_2|, |n_1|, |n_3|\}, & \text{if } sgn(n_1)=-sgn(n_2)=sgn(n_3), \\ max\{|n_2|, |n_1|+|n_3|\}, & \text{if } sgn(n_1)=sgn(n_2)=-sgn(n_3), \\ max\{|n_2|+|n_3|, |n_1|+|n_3|\}, & \text{if } -sgn(n_1)=sgn(n_2)=sgn(n_3). \end{cases}$$

For $m_2(n_4, n_5, n_6)$, we have:
$$m_2(n_4,n_5,n_6) = \begin{cases} |n_4|+|n_5|+|n_6|, & \text{if } sgn(n_4)=sgn(n_5)=sgn(n_6), \\ max\{|n_5|, |n_4|+|n_6|\}, & \text{if } sgn(n_4)=-sgn(n_5)=sgn(n_6), \\ max\{|n_4|+|n_5|, |n_4|+|n_6|\}, & \text{if } sgn(n_4)=sgn(n_5)=-sgn(n_6), \\ max\{|n_4|+|n_6|, |n_5|+|n_6|\}, & \text{if } -sgn(n_4)=sgn(n_5)=sgn(n_6). \end{cases}$$

Thus, we can conclude
$$|n_1|, |n_2|, |n_3|, |n_2|+|n_3| \leq m_1(n),$$
$$|n_i|, |n_i+n_j|, |n_4|+|n_5|+|n_6| \leq m_2(n).$$
where $i$ and $j$ be elements of the set $\{4,5,6\}$, where $i$ is not equal to $j$. We analyze the solution of the given differential equation
$$n'(t) = Qn(t), \quad (23)$$
according to the definition of the norm $\|n\| = max\{m_1, m_2\}$, we address multiple cases.

**Case 1.** When $m_1 > m_2$, $n_1, n_2, n_3 > 0$ and $|n_1| > |n_2 + n_3|$, then we have
$$\|n\| = |n_1| = n_1,$$
hence,

$$D_+\|n\| = n_1'$$
$$= Q_{11}n_1 + Q_{12}n_2 + Q_{13}n_3 + Q_{14}n_4 + Q_{15}n_5 + Q_{16}n_6$$
$$= \left(a - b + a_{11} + a_{22} - \frac{\beta_1 I_u + \beta_2 P}{N}\frac{S}{I_u} - a_{33}\right)n_1 + \left(-\frac{\beta_1 S}{N} + a\right)n_2 + \left[\left(-\frac{\beta_2 S}{N} + a\right)\frac{P}{I_u}\right]n_4$$
$$\leq \left(a - b + a_{11} + a_{22} - \frac{\beta_1 I_u + \beta_2 P}{N}\frac{S}{I_u} - a_{33}\right)|n_1| + \left(-\frac{\beta_1 S}{N} + a\right)|n_2| + \left[\left(-\frac{\beta_2 S}{N} + a\right)\frac{P}{I_u}\right]|n_4|$$
$$\leq \left(a - b + a_{11} + a_{22} - \frac{\beta_1 I_u + \beta_2 P}{N}\frac{S}{I_u} - a_{33} - \frac{\beta_1 S}{N} + a\right)|n_1| + \left[\left(-\frac{\beta_2 S}{N} + a\right)\frac{P}{I_u}\right]|n_4|,$$

where $Q_{ij}(i, j \in 1, 2, ..., 6)$ represents the element in matrix $Q$. And since $|n_4| \leq m_2 < |n_1| = \|n\|$, then

$$D_+\|n\| \le \left(2a - b + a_{11} + a_{22} - a_{33} - 2\frac{\beta_1 S}{N} - \frac{\beta_2 S}{N}\frac{P}{I_u} - \frac{\beta_2 S}{N}\frac{P}{I_u} + a\frac{P}{I_u}\right)\|n\|$$

$$= \left(2a - b + a_{11} + a_{22} - a_{33} - 2b\frac{P}{I_u} + a\frac{P}{I_u}\right)\|n\|$$

$$= \left(a(2+\frac{P}{I_u}) - b(1+2\frac{P}{I_u}) + \theta - (\omega + \eta + \mu)\right)\|n\|$$

**Case 2.** When $m_1 > m_2$, $n_1, n_2, n_3 > 0$ and $|n_1| < |n_2 + n_3|$, then we have

$$\|n\| = |n_2| + |n_3| = n_2 + n_3$$

$$D_+\|n\| = n_2' + n_3'$$
$$= Q_{21}n_1 + Q_{22}n_2 + Q_{23}n_3 + Q_{24}n_4 + Q_{25}n_5 + Q_{26}n_6 + Q_{31}n_1 + Q_{32}n_2 + Q_{33}n_3 + Q_{34}n_4 + Q_{35}n_5 + Q_{36}n_6$$
$$= bn_1 + \left[\frac{\beta_1 S}{N} + a_{11} + a_{33} + \omega - \frac{\beta_1 I_u + \beta_2 P}{N}\frac{S}{I_u} - a_{33}\right]n_2 + \left(\eta + \frac{\beta_1 S}{N} - b + a_{22} + a_{33} - \frac{\beta_1 I_u + \beta_2 P}{N}\frac{S}{I_u} - a_{33}\right)n_3$$
$$+ \left[(\frac{\beta_2 S}{N} - a)\frac{P}{I_u}\right]n_4 + \left[(\frac{\beta_2 S}{N} - a)\frac{P}{I_u}\right]n_5 + \left[(\frac{\beta_2 S}{N} - a)\frac{P}{I_u}\right]n_6$$
$$= bn_1 + \left(\frac{\beta_1 S}{N} + a_{11} + \omega - b\frac{S}{I_u}\right)n_2 + \left(\eta + \frac{\beta_1 S}{N} - b + a_{22} - b\frac{S}{I_u}\right)n_3 + \left[(\frac{\beta_2 S}{N} - a)\frac{P}{I_u}\right](n_4 + n_5 + n_6)$$
$$\le b|n_1| + \left(\frac{\beta_1 S}{N} + a_{11} + \omega - b\frac{S}{I_u}\right)|n_2| + \left(\eta + \frac{\beta_1 S}{N} - b + a_{22} - b\frac{S}{I_u}\right)|n_3| + \left[(\frac{\beta_2 S}{N} - a)\frac{P}{I_u}\right]|n_4 + n_5 + n_6|,$$

and since $|n_4 + n_5 + n_6| \le m_2 < |n_2| + |n_3| = \|n\|$, then

$$D_+\|n\| \le b|n_1| + \left(\frac{\beta_1 S}{N} + a_{11} + \omega - b\frac{S}{I_u}\right)|n_2| + \left(\eta + \frac{\beta_1 S}{N} - b + a_{22} - b\frac{S}{I_u}\right)|n_3| + \left[(\frac{\beta_2 S}{N} - a)\frac{P}{I_u}\right](|n_2| + |n_3|)$$

$$\le b|n_1| + \left(\frac{\beta_1 S}{N} + a_{11} + \omega - b\frac{S}{I_u} + (\frac{\beta_2 S}{N} - a)\frac{P}{I_u}\right)|n_2| + \left(\eta + \frac{\beta_1 S}{N} - b + a_{22} - b\frac{S}{I_u}(\frac{\beta_2 S}{N} - a)\frac{P}{I_u}\right)|n_3|$$

$$= b|n_1| + \left(a_{11} + \omega - a\frac{P}{I_u}\right)|n_2| + \left(\eta - b + a_{22} - a\frac{P}{I_u}\right)|n_3|$$

$$= b|n_1| + \left(-\mu - a\frac{P}{I_u}\right)|n_2| + \left(-b - \mu - a\frac{P}{I_u}\right)|n_3| \le b|n_1| + \left(-\mu - a\frac{P}{I_u}\right)(|n_2| + |n_3|)$$

$$\le \left(b - \mu - a\frac{P}{I_u}\right)(|n_2| + |n_3|) = \left(b - \mu - a\frac{P}{I_u}\right)\|n\|.$$

For the sake of simplicity, according to the approach used to estimate the measure $\bar{\mu}$, the analysis of the remaining cases can be omitted. This proves that the EE $E^*$ of Model (1) is GAS.

We selected a set of parameters to plot the stability diagram of the EE, thereby verifying the theoretical analysis above. The stability diagram of the EE is presented in Supplementary Note 3.

### 3.4 Forward Bifurcation

**Theorem 7.** If $R_0 = 1$ and $\theta\eta < \omega(\mu + \omega)$, then Model (1) undergoes a forward bifurcation at the critical parameter value $\beta_1 = \beta_1^*$.

**Proof** By applying the center manifold theory, we revisit Model (1) and set $\beta_1 = \beta_1^*$ as the bifurcation parameter. Let $V = x_1$, $S = x_2$, $I_u = x_3$, $P = x_4$, $C = x_5$, $R = x_6$ and define $x = (x_1, x_2, x_3, x_4, x_5, x_6)^T$. Reformulate Model (1) in the form $dx/dt = f(x)$, where $f = (f_1, f_2, f_3, f_4, f_5, f_6)^T$. Consequently, Model (1) is transformed into

$$\begin{cases} \dfrac{dx_1}{dt} := f_1 = u\Lambda + \eta x_2 - (\omega + \mu)x_1, \\ \dfrac{dx_2}{dt} := f_2 = (1-u)\Lambda + \omega x_1 - \dfrac{\beta_1 x_3 + \beta_2 x_4}{N} x_2 - (\eta + \mu)x_2, \\ \dfrac{dx_3}{dt} := f_3 = \dfrac{\beta_1 x_3 + \beta_2 x_4}{N} x_2 - (\theta + \mu)x_3, \\ \dfrac{dx_4}{dt} := f_4 = \theta\phi x_3 - (\varepsilon + \tau + \mu)x_4. \\ \dfrac{dx_5}{dt} := f_5 = \theta\sigma x_3 + \tau x_4 - (d + \mu)x_5, \\ \dfrac{dx_6}{dt} := f_6 = \theta(1 - \phi - \sigma)x_3 + \varepsilon x_4 - \mu x_6. \end{cases} \quad (24)$$

If $R_0 = 1$, then we have

$$\beta_1 = \beta_1^* = \dfrac{(\theta + \mu)(\mu + \eta + \omega)}{(\omega + \mu - \mu u)} - \dfrac{\beta_2 \theta \phi}{(\varepsilon + \tau + \mu)}.$$

The Jacobian matrix of Model (24) at $E_0$ and $\beta_1$ is

$$J(E_0, \beta_1) = \begin{bmatrix} -(\omega + \mu) & \eta & 0 & 0 & 0 & 0 \\ \omega & -(\eta + \mu) & \dfrac{\beta_2 \theta \phi(\omega + \mu - \mu u)}{(\varepsilon + \tau + \mu)(\mu + \eta + \omega)} - (\theta + \mu) & -\dfrac{\beta_2(\omega + \mu - \mu u)}{(\mu + \eta + \omega)} & 0 & 0 \\ 0 & 0 & -\dfrac{\beta_2 \theta \phi(\omega + \mu - \mu u)}{(\varepsilon + \tau + \mu)(\mu + \eta + \omega)} & \dfrac{\beta_2(\omega + \mu - \mu u)}{(\mu + \eta + \omega)} & 0 & 0 \\ 0 & 0 & \theta\phi & -(\varepsilon + \tau + \mu) & 0 & 0 \\ 0 & 0 & \theta\sigma & \tau & -(d + \mu) & 0 \\ 0 & 0 & \theta(1 - \phi - \sigma) & \varepsilon & 0 & -\mu \end{bmatrix},$$

(25)

the characteristic equation of the Eq. (25) is

$$(\lambda + \mu)(\lambda + \eta + \omega + \mu)\lambda\left(\lambda + \varepsilon + \tau + \mu + \dfrac{\beta_2 \theta \phi(\omega + \mu - \mu u)}{(\varepsilon + \tau + \mu)(\mu + \eta + \omega)}\right)(\lambda + d + \mu)(\lambda + \mu) = 0,$$

it can be easily obtained that its characteristic values

$$\lambda_1 = 0, \lambda_2 = \lambda_3 = -\mu, \lambda_4 = -(d + \mu),$$

$$\lambda_5 = -(\eta + \omega + \mu), \lambda_6 = -\left(\varepsilon + \tau + \mu + \dfrac{\beta_2 \theta \phi(\omega + \mu - \mu u)}{(\varepsilon + \tau + \mu)(\mu + \eta + \omega)}\right). \quad (26)$$

Eq.(26) demonstrates the suitability of using the center manifold theory to explore the dynamic properties around $\beta_1 = \beta_1^*$. Next, we compute and present the right eigenvalue $w = (\omega_1, \omega_2, \omega_3, \omega_4, \omega_5, \omega_6)^T$ of Eq. (25), where

$$\omega_1 = \dfrac{\eta}{\omega + \mu}\omega_2, \quad \omega_2 = \dfrac{(\theta + \mu)(\omega + \mu)}{\omega\eta - (\eta + \omega)(\omega + \mu)}\omega_3, \quad \omega_4 = \dfrac{\theta\phi}{\varepsilon + \tau + \mu}\omega_3,$$

$$\omega_5 = \dfrac{1}{d + \mu}(\theta\sigma + \dfrac{\tau\theta\phi}{\varepsilon + \tau + \mu})\omega_3, \quad \omega_6 = \dfrac{1}{\mu}[\theta(1 - \phi - \sigma) + \dfrac{\varepsilon\theta\phi}{\varepsilon + \tau + \mu}]\omega_3. \quad (27)$$

The left eigenvector of the Eq. (25) is $v = (v_1, v_2, v_3, v_4, v_5, v_6)^T$, where

$$v_1 = v_2 = v_5 = v_6 = 0, \quad v_3 = 1, \quad v_4 = \dfrac{\beta_2(\omega + \mu - \mu u)}{(\varepsilon + \tau + \mu)(\mu + \eta + \omega)}. \quad (28)$$

Using the property $v \times w = 1$, we can deduce that

$$\omega_2 = -\frac{(\omega+\mu)(\theta+\mu)(\varepsilon+\tau+\mu)}{(\omega^2+\mu\eta+\mu\omega)[(\varepsilon+\tau+\mu)+R_p(\theta+\mu)]}. \tag{29}$$

From Eq. 27, 28, and 29, we can obtain

$$\begin{aligned}&\omega_1<0, \omega_2<0, \omega_3>0, \omega_4>0, \omega_5>0, \omega_6>0,\\ & v_1=v_2=v_5=v_6=0,\ v_3>0,\ v_4>0.\end{aligned} \tag{30}$$

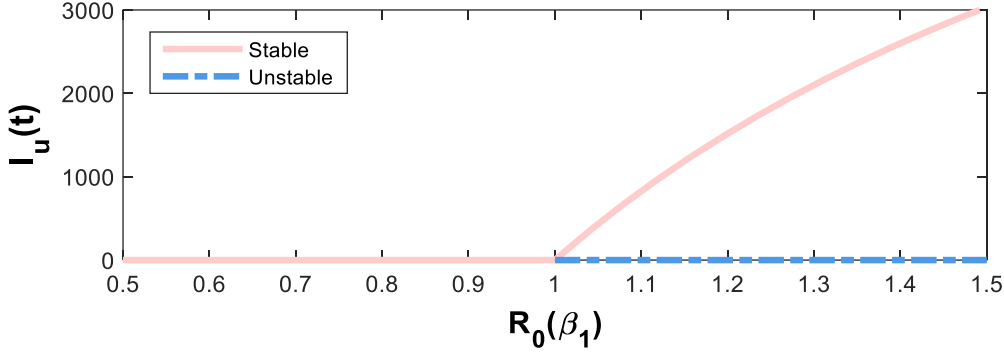

**Fig. 2** The forward bifurcation diagram

Then, the coefficients $a$ and $b$ are computed as follows,

$$a = \sum_{k,i,j=1}^{6} v_k \omega_i \omega_j \frac{\partial^2 f_k}{\partial x_i \partial x_j}(E_0, \beta_1^*)$$

$$= 2v_3\omega_1\omega_3 \frac{\partial^2 f_3}{\partial x_1 \partial x_3}(E_0,\beta_1^*) + 2v_3\omega_1\omega_4 \frac{\partial^2 f_3}{\partial x_1 \partial x_4}(E_0,\beta_1^*) + 2v_3\omega_2\omega_3 \frac{\partial^2 f_3}{\partial x_2 \partial x_3}(E_0,\beta_1^*) + 2v_3\omega_2\omega_4 \frac{\partial^2 f_3}{\partial x_2 \partial x_4}(E_0,\beta_1^*)$$

$$+ v_3\omega_3\omega_3 \frac{\partial^2 f_3}{\partial x_3 \partial x_3}(E_0,\beta_1^*) + 2v_3\omega_3\omega_4 \frac{\partial^2 f_3}{\partial x_3 \partial x_4}(E_0,\beta_1^*) + v_3\omega_4\omega_4 \frac{\partial^2 f_4}{\partial x_4 \partial x_4}(E_0,\beta_1^*)$$

$$= -2\omega_1\omega_3 \frac{\beta_1 S_0}{N^2} - 2\omega_1\omega_4 \frac{\beta_2 S_0}{N^2} + 2\omega_2\omega_3 \frac{\beta_1 V_0}{N^2} + 2\omega_2\omega_4 \frac{\beta_2 V_0}{N^2} - 2\omega_3^2 \frac{\beta_1 S_0}{N^2} - 2\omega_3\omega_4 \frac{\beta_2 S_0}{N^2} - 2\omega_4^2 \frac{\beta_2 S_0}{N^2}$$

$$= -2\omega_3 \frac{\beta_1 S_0}{N^2}(\omega_1+\omega_3) - 2\omega_4 \frac{\beta_2 S_0}{N^2}(\omega_1+\omega_3) + 2\omega_2\omega_3 \frac{\beta_1 V_0}{N^2} + 2\omega_2\omega_4 \frac{\beta_2 V_0}{N^2} - 2\omega_4^2 \frac{\beta_2 S_0}{N^2}.$$

According to Eq. 27, we have $\omega_1+\omega_3 = \frac{\theta\eta-\omega(\mu+\omega)}{(\omega+\mu)(\theta+\mu)}\omega_2$. When $\theta\eta<\omega(\mu+\omega)$, according to inequality (30), we have $\omega_1+\omega_3>0$. Based on this, we ensure that each term in the coefficients $a$ is negative, i.e., $a<0$..

$$b = \sum_{k,i,j=1}^{6} v_k \omega_i \frac{\partial^2 f_k}{\partial x_i \partial \beta_1}(E_0,\beta_1^*) = v_3\omega_3 \frac{\partial^2 f_3}{\partial x_3 \partial \beta_1}(E_0,\beta_1^*) = \frac{\omega+\mu-\mu u}{\mu+\eta+\omega}\omega_3 > 0.$$

Based on the Theorem 4.1 proposed by Castillo-Chavez and Song [36], we can deduce that Model (1) exhibits a forward bifurcation. We select the following set of parameters to verify the existence of the forward bifurcation. $\Lambda = 1000$, $u = 0.01$, $\sigma = 0.01$, $d = 0.00001$, $\eta = 0.01$, $\omega = 0.05$, $\beta_1 = 0.2$, $\beta_2 = 0.1$, $\phi = 0.1$, $\varepsilon = 0.8$, $\tau = 0.001$, $\theta = 0.01687$, $\mu = 0.003$. Fig. 2 shows the forward bifurcation diagram.

## 4 Cervical Cancer in China: A Comprehensive Estimate

In recent years, as China's industrialization and urbanization have deepened, residents' lifestyles have also changed, leading to an increased risk of HPV infection among women. Consequently, the incidence of cervical cancer is rising and showing a trend toward younger age groups. Therefore, accurately estimating the related parameters in the progression of cervical cancer has become particularly important.

We obtain the number of new cervical cancer cases and deaths in mainland China from 1990 to 2021 from the Global Burden of Disease Collaboration Network (Institute for Health Metrics and Evaluation, IHME) [37]. IHME is an independent global health research institution affiliated with the University of Washington, dedicated to providing high-quality health data and assessments of global health trends. Its Global Burden of Disease (GBD) research is widely used to evaluate health risks and disease burdens in different regions.

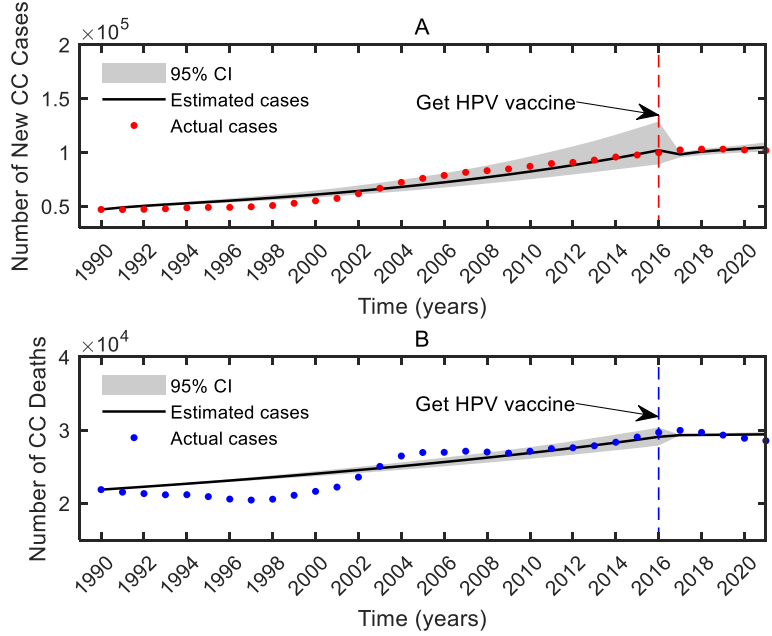

**Fig. 3** Fitting results for the number of new cervical cancer cases and deaths among women from 1990 to 2021. The red and blue dashed lines in subfigures A and B represent the introduction of the HPV vaccine in China in 2016, incorporating the vaccine factor into the fitting process.

Based on existing data and literature, we employ the Markov Chain Monte Carlo (MCMC) algorithm for parameter estimation. We first present the fixed parameter values and initial values for the state variables that are known or can be computed:

(i) From the data provided by IHME [37], we know that in 1990, the number of new cervical cancer cases and the number of deaths among women in mainland China are 47,052 and 21,882, respectively. Thus, we can establish the initial values of the state variables.

(ii) Based on the National Bureau of Statistics of China [38], the total population aged 15-64 is 763.06 million, with a population growth rate of 14.39‰ and a female proportion of 48.48%. Since cervical cancer can only affect women, we assume that the population recruitment rate $\Lambda \approx 5,323,349$.

(iii) Based on the data provided by the National Bureau of Statistics of China [38], the natural mortality rate is 1/70.49, where 70.49 is the average lifespan of women in 1990.

We use the MCMC algorithm to estimate the unknown parameters and the initial values of the state variables. This algorithm is an improvement on MCMC introduced by Haario et al., and further details can be found in reference [39]. We conduct 200,000 iterations using this algorithm, with the first 190,000 iterations designated as a burn-in period. As a result, we estimate the mean, standard deviation of the parameters, as detailed in Table 1 and Table S1. The fitting results are presented in Fig. 3.

Fig. 3 indicates that when the vaccine is introduced in 2016, there is a decline in the growth of both new cervical cancer cases and associated deaths among women. Based on the parameters estimated using MCMC, we obtain the basic reproduction numbers before and after the HPV vaccine is launched in China as 1.5026 (95% CI: 1.4051-1.6002) and 1.0726 (95% CI: 0.9384-1.2067), respectively (see Fig. 4 and Fig. 6). This indicates that HPV continues to circulate in China, but the basic reproduction number decreases by 28.62% after the vaccine is launched, suggesting that the vaccine significantly reduces the transmission capability of HPV.

**Table 1** Parameter values and initial values of the model without vaccination, with 95% confidence intervals

| Parameter | Mean value | Std | Source |
|---|---|---|---|
| $\Lambda$ | 5,323,349 | – | (ii) |
| $\mu$ | 1/70.47 | – | (iii) |
| $\beta_1$ | 0.18447 | 0.00953 | MCMC |
| $\beta_2$ | 0.10936 | 0.05445 | MCMC |
| $\sigma$ | 0.01637 | 0.00468 | MCMC |
| $\theta$ | 0.11071 | 0.00852 | MCMC |
| $\varphi$ | 0.16818 | 0.08756 | MCMC |
| $\varepsilon$ | 0.72317 | 0.12383 | MCMC |
| $\tau$ | 0.01346 | 0.00321 | MCMC |
| $d$ | 0.00402 | 0.00030 | MCMC |
| $S(0)$ | 3.41E+08 | 21,103,939 | MCMC |
| $I_u(0)$ | 84.6941E+04 | 230981.5 | MCMC |
| $P(0)$ | 13.3232E+04 | 33101.69 | MCMC |
| $C(0)$ | 4.7052E+04 | – | (i) |
| $R(0)$ | 2790.67E+04 | – | MCMC |

In addition, we use the Pearson correlation coefficient to measure the relationship between cervical cancer screening rates and the basic reproduction number, which is a statistic used to quantify the linear relationship between two variables. Fig. 5 and Fig. 7 demonstrate a negative correlation between the basic reproduction number and cervical cancer screening rates. This negative correlation, regardless of whether the HPV vaccine is launched, is largely contributed by patients who are unaware of their HPV infection. This finding aligns with real-world scenarios, as patients who are unaware of their HPV infection often lack sufficient understanding of the importance of screening and tend to ignore its role, leading to increased virus transmission. When screening rates are increased for those unaware of their HPV infection, the basic reproduction number decreases, significantly lowering the risk of virus transmission.

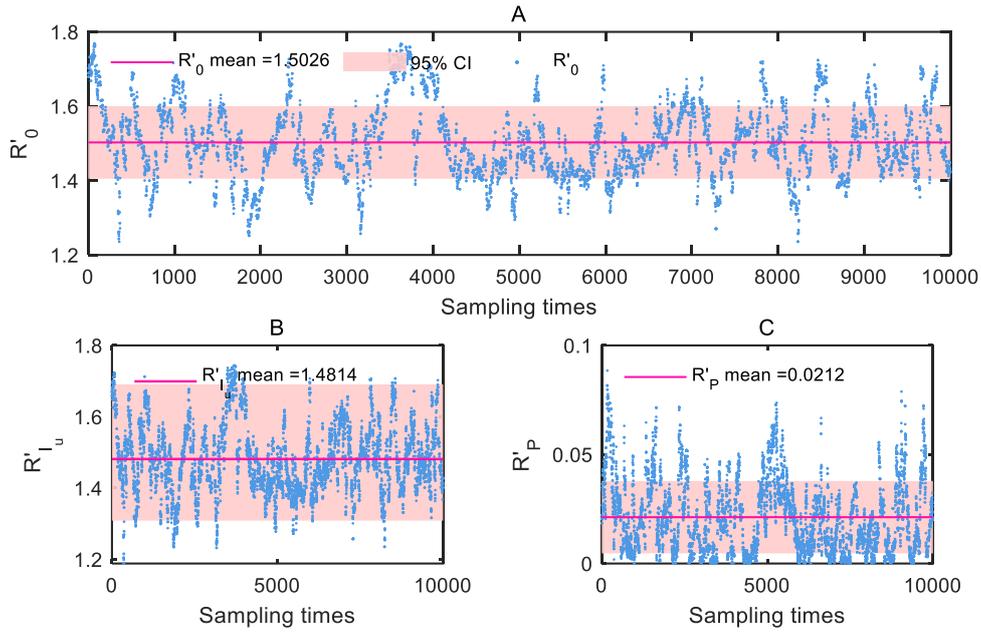

**Fig. 4** The basic reproduction numbers (before the HPV vaccine is launched). **A** $R'_0$ ; **B** $R'_{I_u}$ ; **C** $R'_P$ These values are calculated from the final 10,000 samples of the MCMC fitting results.

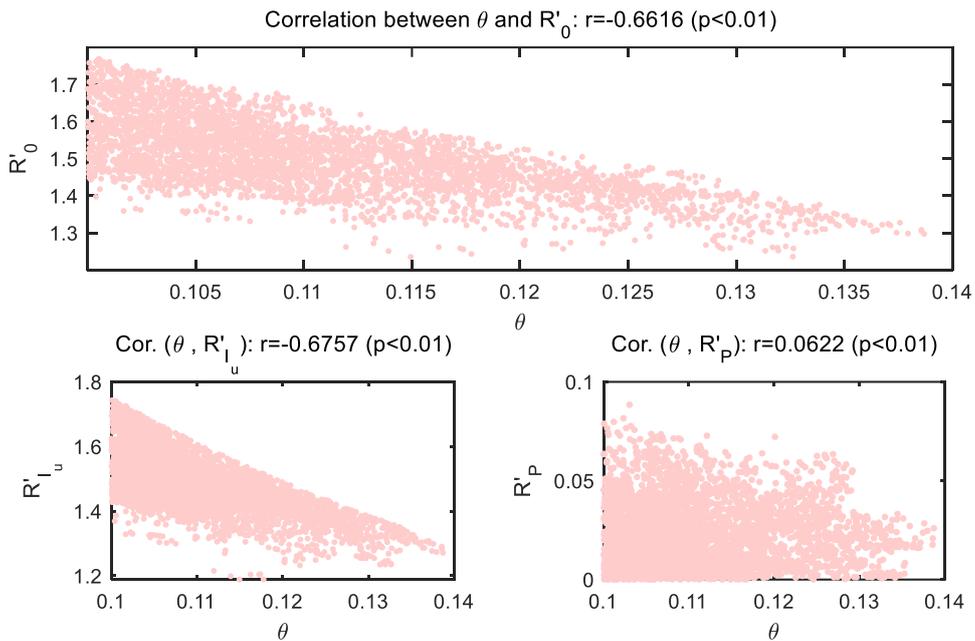

**Fig. 5** Correlation between the basic reproductive number and screening rate (before the HPV vaccine is launched). **A**: Correlation between $\theta$ and $R'_0$ ; **B**: Correlation between $\theta$ and $R'_{I_u}$ ; **C**: Correlation between $\theta$ with $R'_P$ . These values are calculated from the last 10,000 samples of the MCMC fitting results.

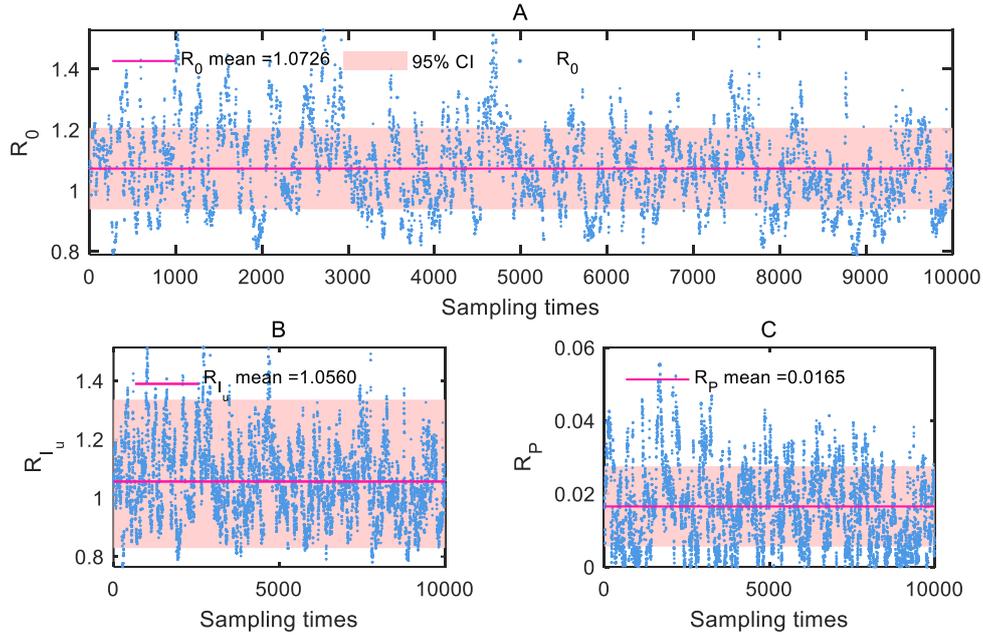

**Fig. 6** The basic reproduction numbers (after the HPV vaccine is launched). **A** $R_0$; **B** $R_{I_u}$; **C** $R_P$. These values are calculated from the final 10,000 samples of the MCMC fitting results.

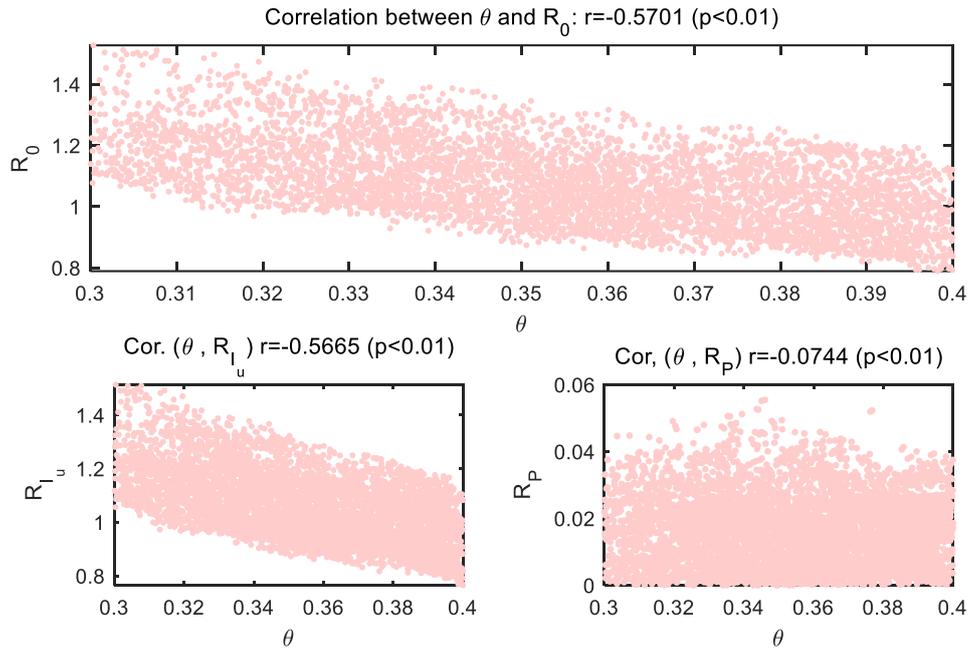

**Fig. 7** Correlation between the basic reproductive number and screening rate (after the HPV vaccine is launched). **A**: Correlation between $\theta$ and $R_0$; **B**: Correlation between $\theta$ and $R_{I_u}$; **C**: Correlation between $\theta$ with $R_P$. These values are calculated from the last 10,000 samples of the MCMC fitting results.

## 4.1 Sensitivity Analysis

In order to reduce the transmission capability of HPV and curb the development of cervical cancer among women in China, we conduct a sensitivity analysis of Model (1) in this section to evaluate the impact of different parameters on disease progression. In the following part, we conduct a sensitivity analysis of $R_0$, using parameter values derived from the MCMC results presented in Table S1.

The analysis in Fig. 8 indicates that $R_0$ is sensitive to changes in transmission rates, with increases in $\beta_1$, $\beta_2$, and $\phi$ resulting in corresponding increases in $R_0$, thereby indicating an elevated risk of disease transmission. Conversely, an increase in the recovery rate $\varepsilon$ is associated with a decrease in $R_0$, suggesting that improved treatment outcomes can significantly curb the spread of the epidemic. Additionally, vaccination strategies are highlighted as effective tools for disease control, as an increase in the vaccination rate for adolescent females $u$ and adult females $\eta$ leads to a notable reduction in $R_0$. This underscores the importance of vaccination programs in reducing the overall disease burden. Similarly, the impact of screening $\theta$ also has a negative effect on $R_0$, indicating that effective screening initiatives play a crucial role in halting the advancement of HPV into cervical cancer.

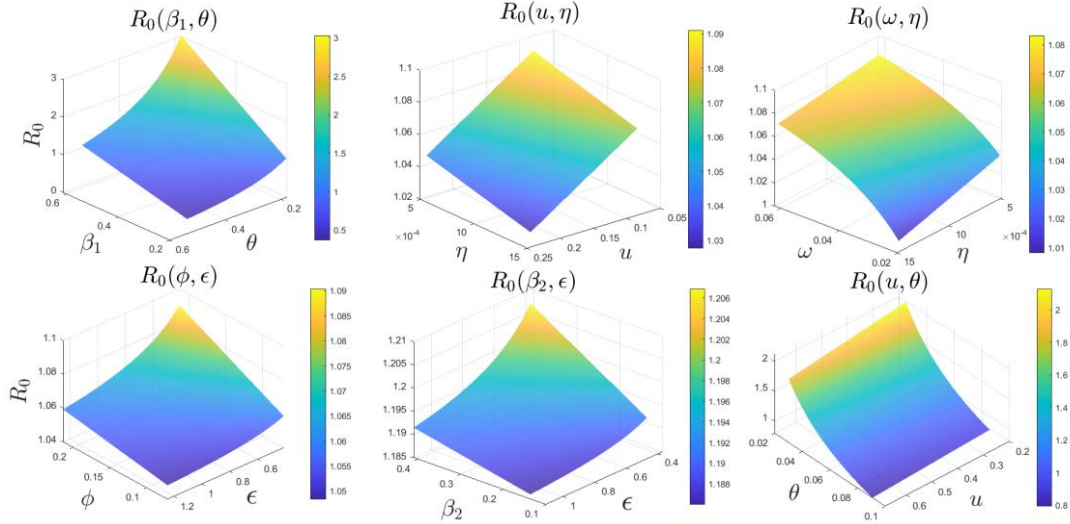

**Fig. 8** Three-dimensional trend chart of $R_0$ with different parameters

We perform a global sensitivity analysis of the compartments in model (1) using Latin Hypercube Sampling (LHS) and Partial Rank Correlation Coefficient (PRCC). LHS divides the parameter space into equal probability intervals and performs random sampling to ensure a uniform distribution of samples. PRCC quantifies the independent effects of various parameters on the model output, helping to assess the relationship between specific parameters and model results. We select a uniform distribution as the prior distribution and perform 1,000 samples. The parameter values listed in Table S1 serve as baseline values, with fluctuations of ±20% applied as input parameters. The sensitivity of different compartments at $t = 4000$ is presented in Fig. 9.

From Fig. 9, we can clearly observe the sensitivity of different parameters to various compartments at $t = 4000$, as well as the positive and negative impacts of these parameters on the compartments. Subfigure **A** indicates that, in addition to the vaccination rate and the recruitment rate of susceptible, the cervical cancer screening rate also positively affects the vaccinated population. This is primarily because cervical cancer screening enhances health awareness and encourages individuals to seek preventive measures, thereby increasing the size of the vaccination cohort. Conversely, the rate of vaccine immunity decline negatively impacts the vaccinated population, as the immune protection provided by vaccination may gradually weaken over time, resulting in vaccinated individuals becoming susceptible. Subfigures **B**, **C**, and **D** demonstrate that the screening rate has a significant negative effect on these compartments, indicating its critical role in reducing the scale of infections. In contrast, the transmission rate of individuals unaware of their HPV infection positively influences these compartments, suggesting that these asymptomatic carriers may unknowingly continue to spread the virus. This hidden transmission not only affects the vaccinated population but also places an additional burden on public health systems.

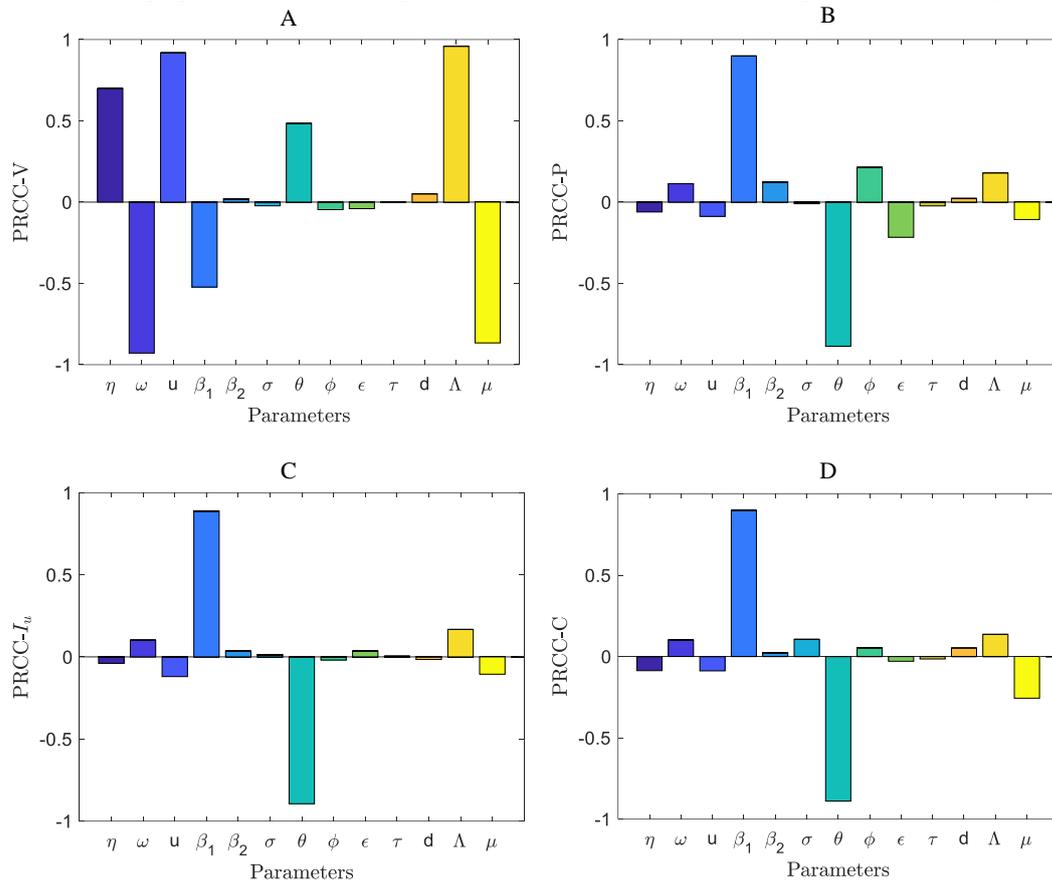

**Fig. 9** Global sensitivity analysis of $V, P, I_u$ and $C$ compartments

## 4.2 Examining the Impact of HPV Vaccine Introduction on Cervical Cancer in China

The HPV vaccine was first approved for use in the United States in 2006, marking it as the world's first vaccine aimed at preventing cervical cancer. Subsequently, multiple countries rapidly followed suit to promote vaccination and reduce the incidence of cervical cancer. China officially introduced the HPV vaccine in 2016, representing a significant step in public health to address the increasingly serious issue of cervical cancer. This study explores the impact of the HPV vaccine's introduction on the scale of cervical cancer in China, measuring the potential reduction in the number of cervical cancer patients and mortality cases attributable to the vaccine. Additionally, we investigate the effects on the infected population had the HPV vaccine been launched one year earlier. The parameters are taken from the values obtained from the MCMC fitting.

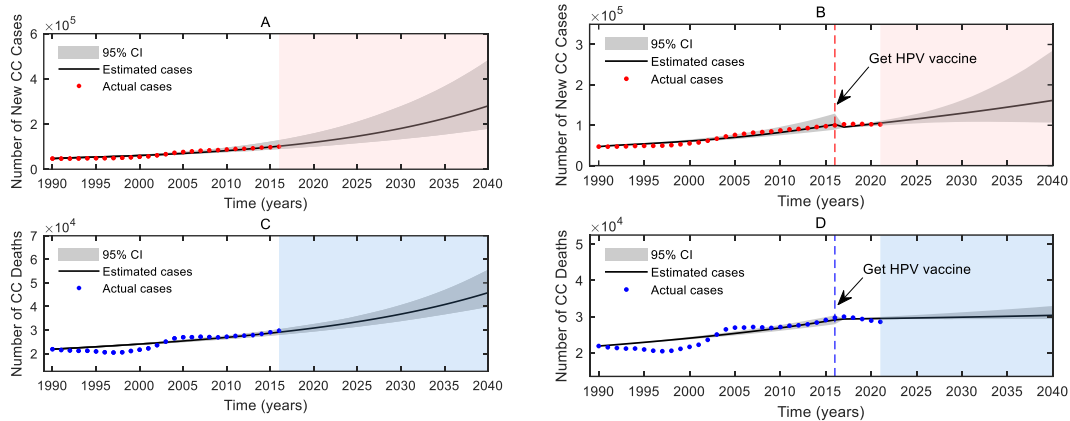

**Fig. 10** Predicted incidence and mortality of cervical cancer in China without and with HPV vaccine introduction. **A C** represent predictions without the HPV vaccine introduction, **B D** represent predictions with the HPV vaccine introduction.

Fig. 10 and Fig. 11 illustrate that following the introduction of the HPV vaccine, the growth curves for both new cases and mortality rates flatten significantly. This indicates that the HPV vaccine has a substantial impact on reducing the incidence of new cases and the number of deaths. By 2040, it is projected that the HPV vaccine will lead to approximately a 42.41% reduction in new cervical cancer cases and a 33.83% decrease in cervical cancer-related deaths. The above highlights that the HPV vaccine has great potential in preventing and controlling cervical cancer. Widespread vaccination can significantly reduce the risk of cervical cancer in women and improve overall public health outcomes.

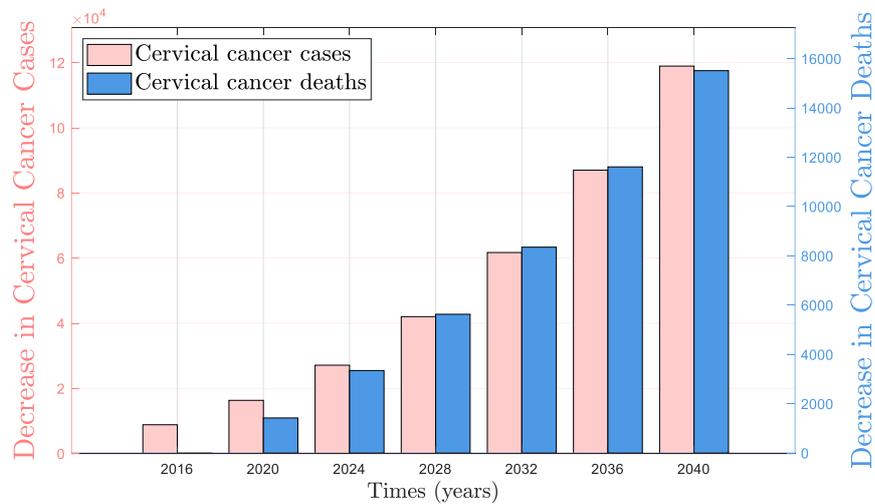

**Fig. 11** Number of new cervical cancer cases and deaths prevented by the introduction of the HPV vaccine by 2040.

Next, we analyze the potential impact of launching the HPV vaccine one year earlier (in 2015) compared to its introduction in 2016 on the number of women spared from developing cervical cancer. As illustrated in Fig. 12, by 2040, it is projected that over 20,000 women are protected from cervical cancer due to the earlier rollout of the vaccine. This data directly reflects the influence of vaccine timing on the incidence of new cervical cancer cases. An earlier implementation of the vaccine enables a greater number of women to avoid cervical cancer compared to the 2016 introduction.

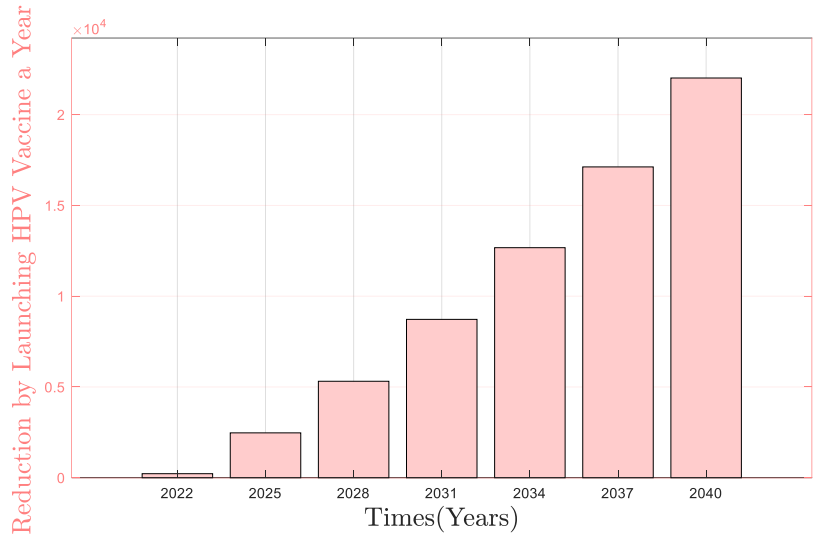

**Fig. 12** Reduction in new cervical cancer cases resulting from the HPV vaccine being launched one year earlier (2015) compared to its introduction in 2016.

## 4.3 Prospective Developments of Cervical Cancer in China Under the 90-70-90 Strategy

In 2020, the WHO proposed the 90-70-90 strategy, aimed at alleviating the burden of cancer through comprehensive vaccination, screening, and treatment initiatives. Subsequently, countries, including China, actively responded to this policy. According to data from 2018-2019, the screening coverage for cervical cancer among women aged 35-64 was 36.8% [8]. Moreover, by the end of 2020, approximately 12.28 million doses of the HPV vaccine had been administered in China, a substantial increase from about 3.42 million doses at the end of 2018 [40]. This indicates that China has made progress in advancing cervical cancer prevention and control efforts.

We quantified the 90-70-90 strategy proposed by the WHO using mathematical methods and integrated the results of this quantification with the established model, as illustrated in Fig. 13. In Fig. 13 **A**, we quantify the first goal, which is to ensure that 90% of girls complete the HPV vaccination before the age of 15. Among the recruits, we assume that 90% of girls have been vaccinated against HPV and enter the vaccinated compartment $V(t)$, while 10% have not been vaccinated and enter the susceptible compartment $S(t)$.

The second goal is quantified in Fig. 13 **B**, where the susceptible population and patients unaware of their HPV infection are considered as the screening population. We assume that, after screening, the susceptible individuals return to the susceptible compartment $S(t)$. It is hypothesized that 70% of those unaware of their HPV infection undergo screening, resulting in three possible outcomes: a diagnosis of precancerous lesions, a diagnosis of invasive cervical cancer, or a normal screening result (the latter may occur due to the possibility that individuals unaware of their infection may not be detectable within a short timeframe or may have cleared the virus through their immune system). We assume that 70% of those unaware of their infection are screened per unit time. Consequently, the transition rate from the susceptible compartment $I_u(t)$ to the recovered compartment $R(t)$ can be expressed as $(1-\phi-\sigma) \times 70\% \times I_u \times 1/\text{unit time}$; the transition rate from $I_u(t)$ to the precancerous lesion compartment $P(t)$ can be expressed as $\phi \times 70\% \times I_u \times 1/\text{unit time}$; and the transition rate from $I_u(t)$ to the cervical cancer compartment $C(t)$ can be expressed as $\sigma \times 70\% \times I_u \times 1/\text{unit time}$.

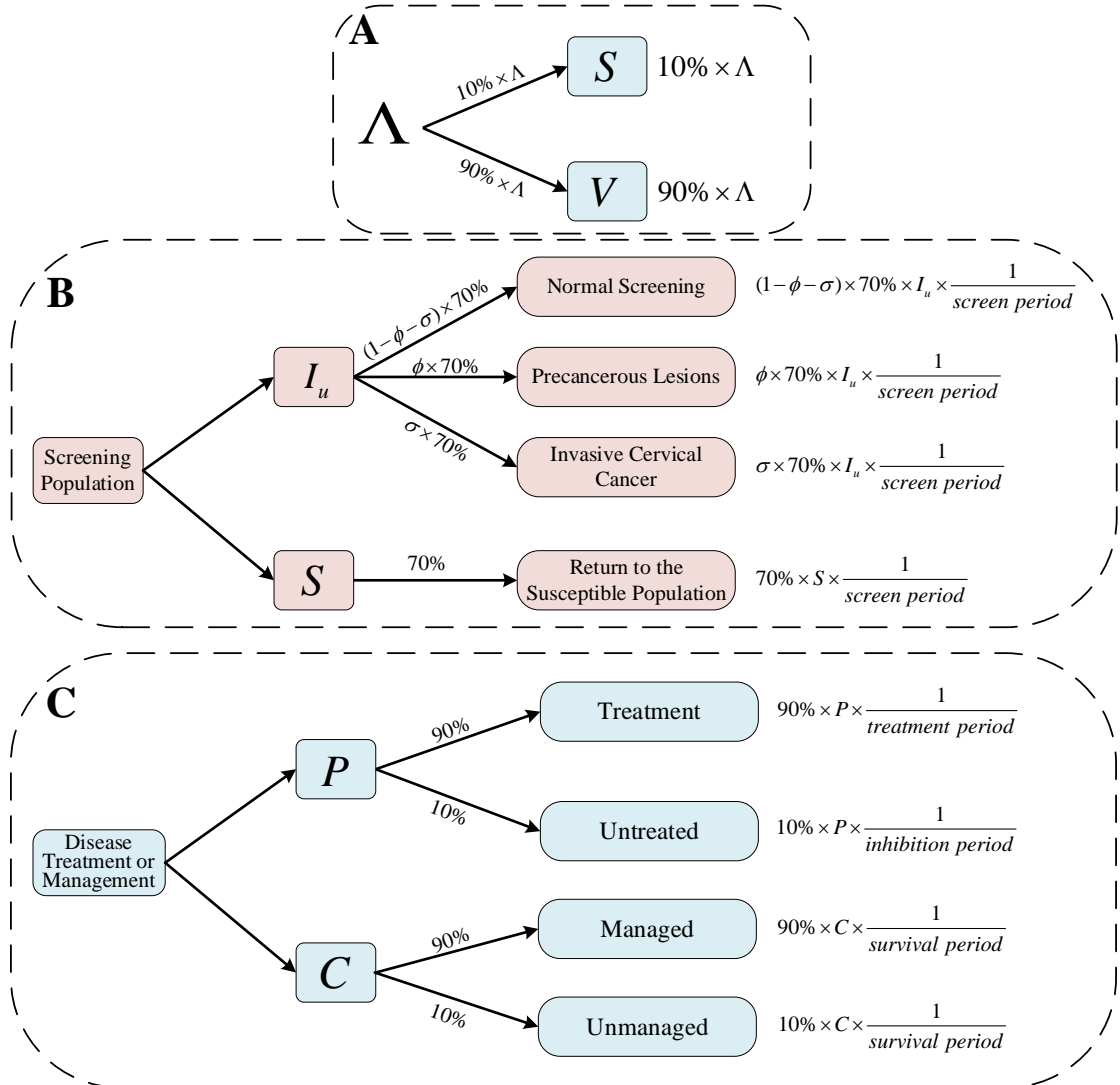

**Fig. 13** Mathematical Quantification of the 90-70-90 Targets and Integration with the Model. **A**: 90% of girls complete HPV vaccination before the age of 15; **B**: 70% of women undergo screening with effective detection methods before the ages of 35 and 45; **C**: 90% of women diagnosed with cervical diseases receive treatment (90% of women with positive precancerous lesions receive treatment, and 90% of invasive cancer cases are managed).

In Fig. 13 **C**, the quantification focuses on ensuring that 90% of women diagnosed with cervical disease receive treatment. It is assumed that 90% of patients with precancerous lesions receive treatment per unit time, while 10% do not. This process can be represented as $90\% \times P \times 1/\text{unit time}$ and $10\% \times P \times 1/\text{unit time}$, respectively. Similarly, it is assumed that 90% of patients with invasive cervical cancer are managed per unit time, while 10% are not managed. This process is represented as $90\% \times C \times 1/\text{unit time}$ and $10\% \times C \times 1/\text{unit time}$, respectively.

According to the above assumptions, we consider when China can achieve the goal of zero new cases of cervical cancer after reaching the 90-70-90 targets by the end of 2030. We analyze the following situations:

**Case 1.** By 2030, 90% of girls complete HPV vaccination before the age of 15.
**Case 2.** By 2030, 70% of women of appropriate age receive cervical cancer screening.
**Case 3.** By 2030, 90% of women diagnosed with cervical disease receive treatment.

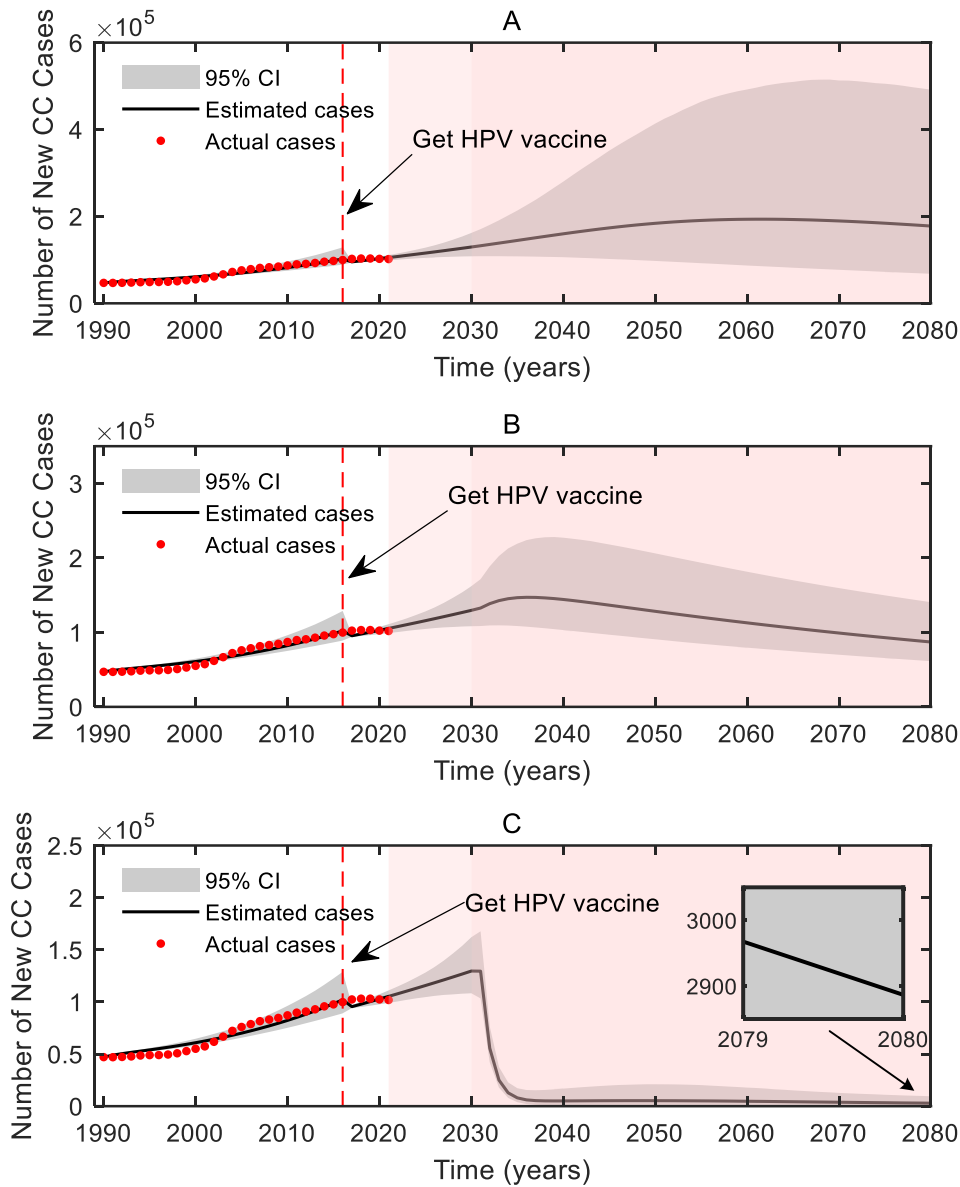

**Fig. 14** Achievement of a single target from the 90-70-90 framework by 2030. **A**: Case 1; **B**: Case 2; **C**: Case 3. The light red area represents the forecast range, while the slightly darker red area indicates the projections following target achievement.

**Case 4.** By 2030, 90% of girls complete HPV vaccination before the age of 15, and 70% of women of appropriate age receive cervical cancer screening.
**Case 5.** By 2030, 90% of girls complete HPV vaccination before the age of 15, and 90% of women diagnosed with cervical disease receive treatment.
**Case 6.** By 2030, 70% of women of appropriate age receive cervical cancer screening, and 90% of women diagnosed with cervical disease receive treatment.
**Case 7.** By 2030, 90% of girls complete HPV vaccination before the age of 15, 70% of women of appropriate age receive cervical cancer screening, and 90% of women diagnosed with cervical disease receive treatment.

Fig. 14 shows that by 2030, achieving only one of the targets of 90-70-90 does not lead to the elimination of new cervical cancer cases in China. In Fig. 14 B, as screening rates increase, many previously undiagnosed cases are identified during the initial phase. This leads to a significant short-term increase in the number of cancer cases. However, as time progresses, the benefits of early screening gradually become evident, and the number of new cervical cancer cases significantly declines. In Fig. 14 C, when 90% of women diagnosed with cervical diseases receive treatment, although it does not eliminate new cervical cancer cases in the long term, it substantially reduces the burden of cervical cancer compared to a scenario without effective intervention.

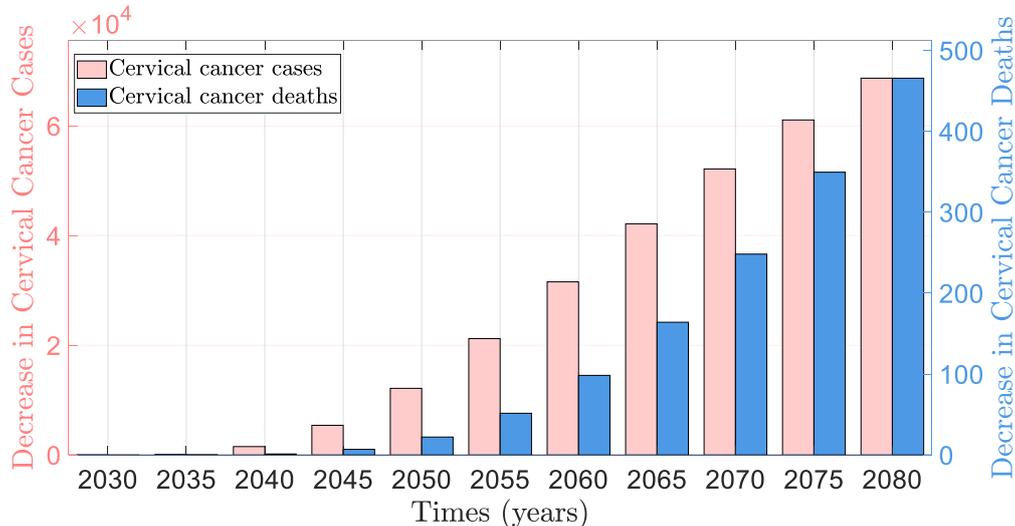

**Fig. 15** Reduction in the number of new cervical cancer cases and deaths in 2030 under Case 1.

From Fig. 15, it is evident that vaccination before the age of 15 reduces the incidence of cervical cancer and associated mortality. It is projected that by 2080, this strategy will prevent 68,710 new cases of cervical cancer and save 465 lives. Following HPV vaccination, there is no immediate decline in the incidence and mortality of cervical cancer because the girls vaccinated before age 15 have not yet reached the age at which the disease typically occurs, which is usually between 30 and 55 years. In recent years, there is a trend of younger women developing cervical cancer, with a small number of cases occurring in those aged 25 and under (i.e., around 2045). Therefore, it is only in the 10 to 15 years after 2030, as these protected individuals reach the age range where cervical cancer may occur, that there will be a significant reduction in cancer cases and mortality.

Fig. 16 shows that even with the achievement of two scenarios outlined in the 90-70-90 target by 2030, specifically case 4 (Fig. 16A) and case 5 (Fig. 16B), this goal remains unattainable. However, when considering case 6 (Fig. 16C), it is projected that by 2061, the incidence of new cervical cancer cases will reach zero. If the 90-70-90 target is fully realized (see Fig. 17), this outcome is expected to be achieved two years earlier, in 2059. This suggests that implementing case 6, which entails 70% screening and 90% treatment rates, represents the most cost-effective option given limited healthcare resources. This conclusion aligns perfectly with the Chinese government's 2023 announcement regarding the Action Plan for Accelerating the Elimination of Cervical Cancer (2023-2030), which aims to ensure that by 2030, the HPV vaccination pilot program continues for eligible girls, the cervical cancer screening rate for eligible women reaches 70%, and the treatment rate for cervical cancer and pre-cancerous lesions reaches 90%.

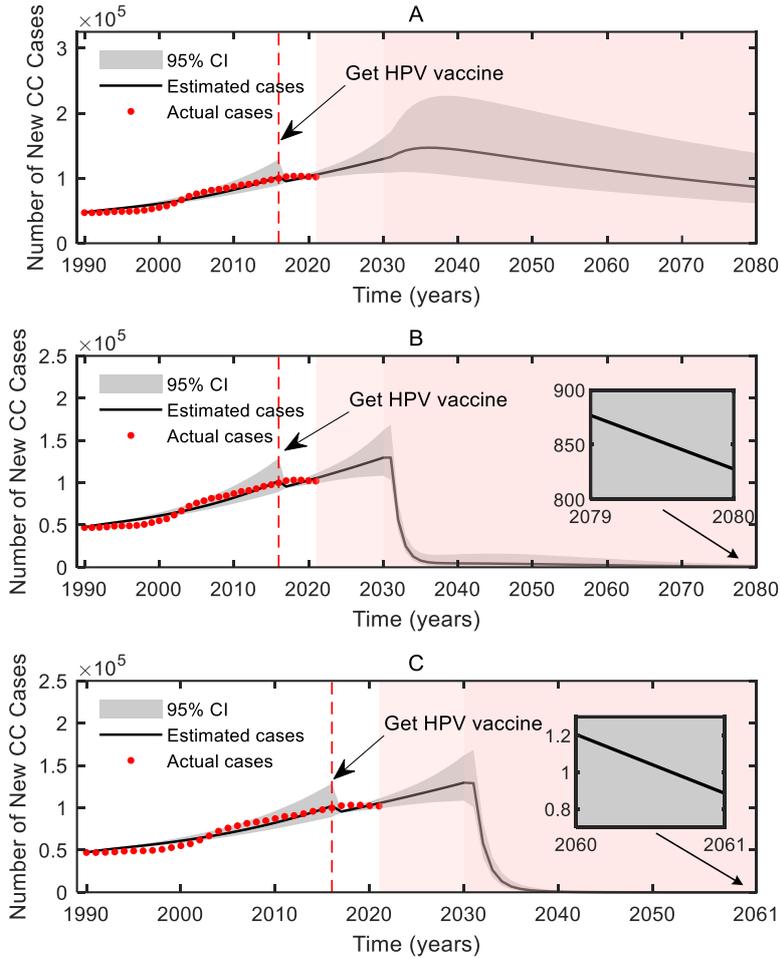

**Fig. 16** Achievement of two targets from the 90-70-90 framework by 2030. **A**: Case 4; **B**: Case 5; **C**: Case 6. The light red area represents the forecast range, while the slightly darker red area indicates the projections following target achievement.

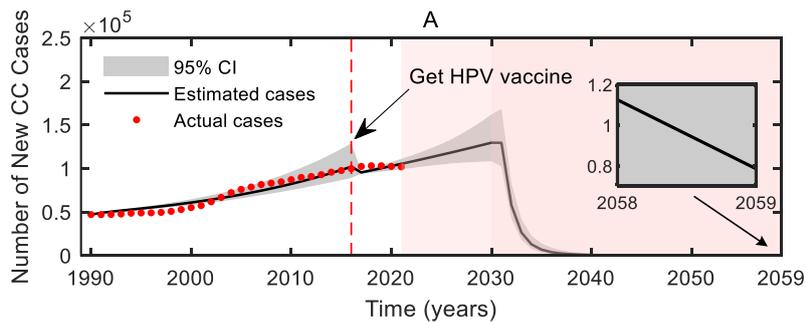

**Fig. 17** Full achievement of the 90-70-90 targets by 2030, i.e., Case 7. The light red area represents the forecast range, while the slightly darker red area indicates the projections following target achievement.

# 5 The Optimal Control Problem

## 5.1 Optimal Control Analysis

In this section, we explore the optimal control problem to determine the best strategy combinations for cervical cancer prevention and screening [41-42]. By applying optimal control theory, we aim to identify how to adjust vaccination rates and screening rates to achieve the maximum reduction in disease burden. This approach seeks to minimize new cases and related deaths within a specified timeframe.

We consider the adult vaccination rates, the vaccination rates of girls under 15, and the screening rates as control variables, denoted as $u_1(t)$, $u_2(t)$ and $u_3(t)$, respectively. We aim to determine the most effective strategy that reduces the health burden of HPV and lowers the total costs of vaccination and screening. The corresponding state system for System (1) is described by:

$$\begin{cases} \dfrac{dV}{dt} = u_2\Lambda + u_1 S - (\omega + \mu)V, \\ \dfrac{dS}{dt} = (1-u_2)\Lambda + \omega V - \dfrac{\beta_1 I_u + \beta_2 P}{N}S - (u_1 + \mu)S, \\ \dfrac{dI_u}{dt} = \dfrac{\beta_1 I_u + \beta_2 P}{N}S - (u_3 + \mu)I_u, \\ \dfrac{dP}{dt} = u_3\phi I_u - (\varepsilon + \tau + \mu)P, \\ \dfrac{dC}{dt} = u_3\sigma I_u + \tau P - (d+\mu)C, \\ \dfrac{dR}{dt} = u_3(1-\phi-\sigma)I_u + \varepsilon P - \mu R, \end{cases} \quad (31)$$

$V(0) = V_0, S(0) = S_0, I_u(0) = I_{u_0}, P(0) = P_0, C(0) = C_0, R(0) = R_0.$

We introduce the following objective function to assess the costs associated with vaccination, the costs of screening, and the health burden associated with patients who are unaware of their HPV infections ($I_u$), those in the precancerous stage ($P$), and patients with invasive cervical cancer ($C$):

$$J(u) = \int_0^T \big(f(t,x,u)\big)dt \\ = \int_0^T \left(\frac{1}{2}A_1 u_1^2(t) + \frac{1}{2}A_2 u_2^2(t) + \frac{1}{2}A_3 u_3^2(t) + B_1 I_u + B_2 P + B_3 C\right)dt, \quad (32)$$

where $u = \{u_1, u_2, u_3\}$. $x$ solves System(31) for the specified control $u$, the constants $A_1, A_2, A_3, B_1, B_2, B_3$ describes the relative impact of control variables or state variables on the objective function value, $T$ is the final time, with a control set defined by

$$U = \{u \mid \text{bounded and Lebesgue measurable on } [0,T]\}, \quad (33)$$

with bounded defined by

$$0 \leq u_i(t) \leq u_i^{\max}, \ i=1,2,3, \ \forall t \in [0,T]. \quad (34)$$

We aim to address the following optimal control problem: Identify an optimal control pair $u^*$ that minimizes

$$J(u^*) = \min\{J(u) \mid u \in U\} \quad (35)$$

while adhering to the system defined by System (31). The first task is to determine whether optimal controls exist and to characterize them.

**Theorem 8** There exists a solution $u^*(t) = (u_1^*(t), u_2^*(t), u_3^*(t))^T$ to the optimal control problem (35).

**Proof** We apply Theorem 4.1 from Kern et al. [43] to prove the existence of $u^*$. We check the following assumptions:
(H1) The set of controls and corresponding state variables is not empty.
(H2) The set of measurable controls is both convex and closed.

(H3) Each term on the right side of the state system (31) is continuous, bounded above by a sum of the bounded control and state, and can be expressed as a linear function of the control variable, with coefficients that depend on time and state.

(H4) The integrand $f(t,x,u)$ of the objective functional is convex.

(H5) There exist constants $c_1 > 0, c_2 > 1, c_3 > 0$ such that the integrand of the objective functional satisfies the inequality $f \geq c_1(|u_1|^2 + |u_2|^2 + |u_3|^2)^{c_2} - c_3$.

It is clear that the set of controls and their corresponding state variables is not empty, and the admissible control set is both convex and closed. The Lipschitz continuity and boundedness of the state system ensure that assumption (H3) holds. Concerning (H4), we apply the approach of Saldaña et al. [18] to demonstrate that (H4) is valid. Notably, since

$$f(t,x,u) \geq \frac{1}{2}A_1 u_1^2(t) + \frac{1}{2}A_2 u_2^2(t) + \frac{1}{2}A_3 u_3^2(t),$$

we can set $c_1 = \min \frac{A_i}{2} (i=1,2,3)$, $c_2 = 2$ and $c_3 = 0$. Therefore, (H5) is satisfied. The proof is complete.

Let $\xi = (\xi_1, \xi_2, \xi_3, \xi_4, \xi_5, \xi_6)^T$. To determine the optimal solution, we define the Hamiltonian function $H$ for the control problem as follows,

$$H = \xi_1\left[u_2\Lambda + u_1 S - (\omega + \mu)V\right] + \xi_2\left[(1-u_2)\Lambda + \omega V - \frac{\beta_1 I_u + \beta_2 P}{N}S - (u_1 + \mu)S\right]$$

$$+ \xi_3\left[\frac{\beta_1 I_u + \beta_2 P}{N}S - (u_3 + \mu)I_u\right] + \xi_4\left[u_3\phi I_u - (\varepsilon + \tau + \mu)P\right] + \xi_5\left[u_3\sigma I_u + \tau P - (d+\mu)C\right]$$

$$+ \xi_6\left[u_3(1-\phi-\sigma)I_u + \varepsilon P - \mu R\right] + \frac{1}{2}A_1 u_1^2(t) + \frac{1}{2}A_2 u_2^2(t) + \frac{1}{2}A_3 u_3^2(t) + B_1 I_u + B_2 P + B_3 C,$$

where $\xi_i (i=1,2,3,4,5,6)$ are the adjoint variables. Applying Pontryagin's Maximum Principle [44], we derive the following result:

$$\frac{d\xi_1}{dt} = -\left[-\xi_1(\mu+\omega) + \xi_2(\omega + \frac{\beta_1 I_u + \beta_2 P}{N^2}S) - \xi_3\frac{\beta_1 I_u + \beta_2 P}{N^2}S\right],$$

$$\frac{d\xi_2}{dt} = -\left[\xi_1 u_1 - \xi_2(\mu + u_1 + \frac{\beta_1 I_u + \beta_2 P}{N} - \frac{\beta_1 I_u + \beta_2 P}{N^2}S) + \xi_3(\frac{\beta_1 I_u + \beta_2 P}{N} - \frac{\beta_1 I_u + \beta_2 P}{N^2}S)\right],$$

$$\frac{d\xi_3}{dt} = -\left[B_1 + \xi_2(\frac{\beta_1 I_u + \beta_2 P}{N^2}S - \frac{\beta_1 S}{N}) - \xi_3(\mu + u_3 + \frac{\beta_1 I_u + \beta_2 P}{N^2}S - \frac{\beta_1 S}{N}) + \xi_4\phi u_3 + \xi_5\sigma u_3 + \xi_6 u_3(1-\phi-\sigma)\right],$$

$$\frac{d\xi_4}{dt} = -\left[B_2 + \xi_2(\frac{\beta_1 I_u + \beta_2 P}{N^2}S - \frac{\beta_2 S}{N}) - \xi_3(\frac{\beta_1 I_u + \beta_2 P}{N^2}S - \frac{\beta_2 S}{N}) - \xi_4(\varepsilon + \mu + \tau) + \xi_5\tau + \xi_6\varepsilon\right],$$

$$\frac{d\xi_5}{dt} = -\left[B_3 + \xi_2\frac{\beta_1 I_u + \beta_2 P}{N^2}S - \xi_3\frac{\beta_1 I_u + \beta_2 P}{N^2}S - \xi_5(d+\mu)\right],$$

$$\frac{d\xi_6}{dt} = -\left[\xi_2\frac{\beta_1 I_u + \beta_2 P}{N^2}S - \xi_3\frac{\beta_1 I_u + \beta_2 P}{N^2}S - \xi_6\mu\right],$$

$\xi_i(T) = 0 \ \forall i = 1,2,\cdots,6$.

Additionally, the optimal controls meet the following equations,

$$u_1^*(t) = \max\left(\min\left(\frac{S(\xi_2 - \xi_1)}{A_1}, u_1^{\max}\right), 0\right),$$

$$u_2^*(t) = \max\left(\min\left(\frac{\Lambda(\xi_2 - \xi_1)}{A_2}, u_2^{\max}\right), 0\right),$$

$$u_3^*(t) = \max\left(\min\left(\frac{I_u[\xi_3 - \xi_4\phi - \xi_5\sigma - \xi_6(1-\phi-\sigma)]}{A_3}, u_3^{\max}\right), 0\right).$$

## 5.2 Numerical Simulations of Optimal Control

Here, we use the forward-backward sweep algorithm to simulate realistic scenarios associated with HPV vaccination and screening, thereby supplementing the previous analytical results. The parameters and initial values used for the numerical simulations are derived from the MCMC fitting results, as shown in Table S1.

HPV vaccination, as well as the clinical management and treatment costs associated with HPV infection, vary between countries. This study primarily examines HPV treatment and vaccination costs in the context of mainland China. In China, the economic burden of administering 2 doses of the HPV vaccine to girls aged 9-14 is approximately 1,200-2,600 CNY. For adults, the nine-valent vaccine is commonly chosen, with a full vaccination cost of about 4,000 CNY [45]. The cost of cervical cancer screening is approximately 500 CNY [46]. According to the study by Li et al. [47], the economic burden for cervical cancer patients and precancerous lesion patients in China is approximately 150,000 CNY and 15,000 CNY, respectively. Therefore, we set the weight parameters as follows: $A_1 = 4000$, $A_2 = 2000$, $A_3 = 500$, $B_2 = 15,000$, and $B_3 = 150,000$. Given the lower economic burden for patients unaware of their HPV infection, we assume $B_1 = A_3$.

In the following simulation, we examine four scenarios to demonstrate the impact of different combinations of control strategies on disease transmission, as outlined below,

$\Pi 1$: Only adult females receive the HPV vaccine $\left(u_2^*(t)\right)$.

$\Pi 2$: Adult females receive the HPV vaccine and undergo screening. $\left(u_2^*(t), u_3^*(t)\right)$.

$\Pi 3$: Girls and adult females receive the HPV vaccination $\left(u_1^*(t), u_2^*(t)\right)$.

$\Pi 4$: Girls and adult females receive the HPV vaccine and undergo screening $\left(u_1^*(t), u_2^*(t), u_3^*(t)\right)$.

Fig. 18 and Fig. 19 respectively show the disease dynamics without control and under four different control strategies ($\Pi 1$ to $\Pi 4$) from 2016 to 2080, along with the corresponding changes in control values. First, Fig. 18 indicates that under strategy $\Pi 1$ (vaccination only for adult females), despite maintaining a high vaccination rate for adult females after 2016, this strategy has limited effectiveness in reducing infection and cancer cases. By around 2030, there is a slight decline in infection rates and cancer burden, but it fails to significantly control HPV transmission over the entire simulation period. This suggests that vaccinating only adult females is insufficient to notably reduce the burden of HPV-related diseases.

In contrast, strategy $\Pi 2$ (vaccination and screening for adult females) combines screening measures, further reducing cases of cervical cancer and precancerous lesions. The control values in Fig. 19 show that the screening rate remains high from 2016 to 2075, but, compared to strategy $\Pi 1$, the combination of these two control measures reduces the time needed to maintain a high vaccination rate until 2052, enabling earlier detection and intervention for HPV-related lesions. This suggests that the combined strategy is more effective than vaccination alone, effectively preventing the long-term spread of HPV.

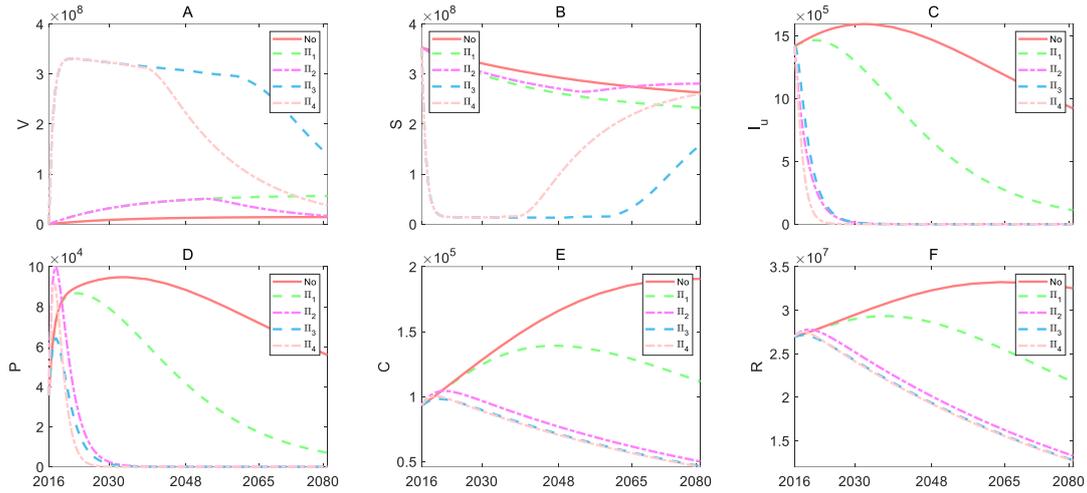

**Fig. 18** Numerical simulations of different compartments under various control strategies. The red line represents no control, while $\Pi_i (i=1,2,3,4)$ corresponds to different control measures, respectively.

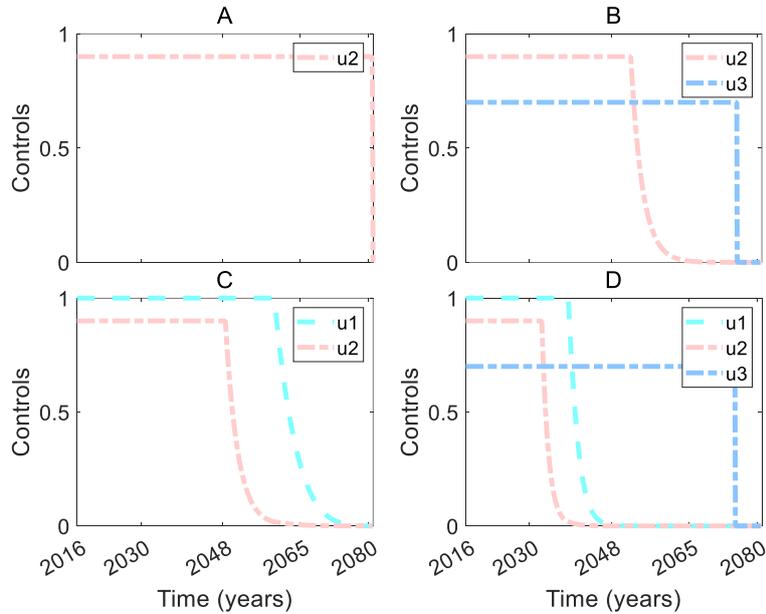

**Fig. 19** Control curves for different strategies. Subfigure **A** corresponds to $\Pi_1$, Subfigure **B** to $\Pi_2$, Subfigure **C** corresponds to $\Pi_3$, and Subfigure **D** corresponds to $\Pi_4$.

In strategy $\Pi 3$ (vaccination for all females), vaccinating all females results in a more pronounced reduction in HPV infections and cancer burden. Fig. 18 shows that this strategy leads to a significant decline in the number of infections and cancer burden after 2030. In Fig.19, the vaccination control value reaches its maximum in the early stages of 2016, indicating the need to maintain a high vaccination rate at the beginning to achieve herd immunity as early as possible. With increasing coverage, the maximum vaccination rate gradually decreases after 2047 and 2058, respectively. Compared to strategy $\Pi 2$, the time required to reach the target outcome is further shortened. This indicates that early vaccination is crucial for controlling HPV transmission. However, while vaccination for females alone is effective, it does not achieve the optimal control outcome.

Strategy $\Pi 4$ (vaccination and screening for all females) is the most effective control combination. Fig. 18 shows that between 2016 and 2030, the number of infections and precancerous lesions significantly decreases, reaching a low level around 2030, nearly achieving complete control of HPV transmission. The control values in Fig. 19 show that under strategy $\Pi_4$, both vaccination and screening rates remain at their highest levels from 2016 to 2030, maximizing the suppression of early HPV transmission risks. After that, both vaccination and screening rates gradually decline, indicating that as immunity coverage increases, the need to maintain high vaccination and screening rates decreases.

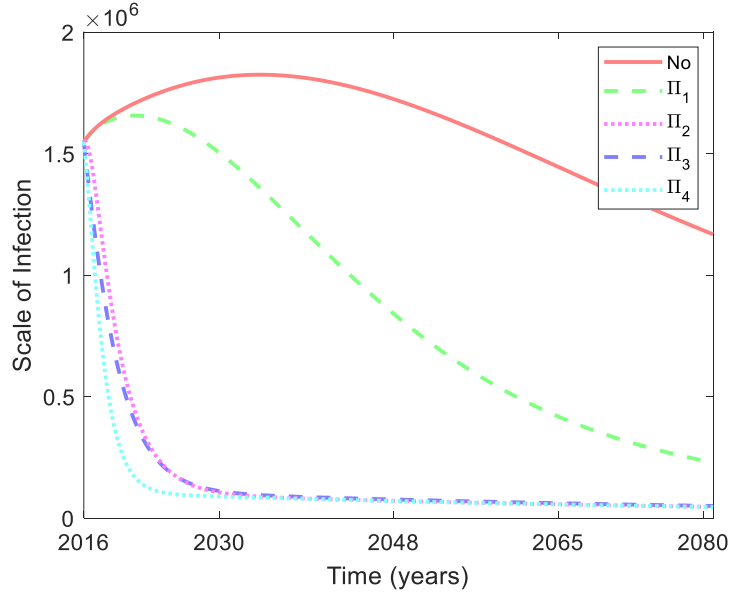

**Fig. 20** Infection scale under various control strategies.

Therefore, the combined strategy of vaccination and screening can minimize the long-term burden of HPV infection and related cancers. Although single measures such as vaccination or screening can reduce some disease burden in the early stages, their effects are insufficient to achieve a significant reduction in the infection scale in a short period (see Fig. 20). In contrast, under resource availability, strategy $\Pi_4$ (vaccination and screening for all females) provides the best long-term control effect, reducing the HPV-related disease burden more quickly and effectively. However, in resource-limited settings, the combination strategy of vaccination and screening for adult females ($\Pi_2$) still offers high cost-effectiveness, enabling partial disease control at a lower cost, thereby reducing the infection scale.

## 6 Conclusion

In this paper, we establish a compartmental model based on the WHO's 90-70-90 targets, incorporating HPV vaccination for girls, adult vaccination catch-up, screening, and treatment. We first validate the model's invariant set and calculate the basic reproduction number using the next-generation matrix approach. Using Routh-Hurwitz criteria, the Lyapunov function, and geometric methods, we prove that both the disease-free and endemic equilibria are locally and globally asymptotically stable. Additionally, we demonstrate a forward bifurcation in the model, which we further verify through numerical simulations.

Based on data from 1990 to 2021 on cervical cancer cases and mortality in China, we employ MCMC algorithms to estimate the unknown parameters and initial values. Given that the HPV vaccine was introduced in mainland China in 2016, Our model estimates a basic reproduction number of 1.5026 (95% CI: 1.4051-1.6002) before vaccine introduction and a reproduction number of 1.0726 (95% CI: 0.9384-1.2067) after vaccine implementation, indicating a 28.62% reduction in basic reproduction number after vaccine introduction. This suggests a significant reduction in transmission rates due to vaccination. Sensitivity analysis shows that screening, vaccination, and catch-up vaccination rates substantially impact the model's outcomes.

We further assess the impact of HPV vaccination on reducing cervical cancer cases in China by comparing scenarios with and without vaccination. By 2040, HPV vaccination is projected to reduce new cervical cancer cases by 42.41% and deaths by 33.83%. If the HPV vaccine had been introduced one year earlier (in 2015), more than 20,000 additional cases would be prevented by 2040, emphasizing the importance of early vaccine introduction.

To quantify the WHO's 90-70-90 strategy, we integrate the target values with our model to evaluate possible outcomes if China achieves these targets by 2030, simulating scenarios with partial and full achievement of the goals. The results show that reaching only one of the targets is insufficient for eliminating cervical cancer. With the attainment of the 70% screening and 90% treatment targets by 2030, zero new cases could be expected by 2061. Full realization of the 90-70-90 targets by 2030 could accelerate this to 2059, suggesting that in resource-limited settings, a strategy focusing on 70% screening and 90% treatment offers the highest cost-effectiveness. This aligns with China's National Cervical Cancer Action Plan.

Finally, through optimal control analysis, we explore the most effective prevention strategy combinations under varying resource conditions. Simulation results indicate that a strategy of universal HPV vaccination and screening substantially reduces HPV infections, cervical cancer burden, and infection scale in the long term. However, in resource-constrained settings, combining vaccination and screening for adult women remains highly cost-effective, controlling the disease burden at a lower cost and consequently reducing the infection scale.

**CRediT authorship contribution statement**

**Hua Liu:** Methodology, Conceptualization, Visualization, Software. **Chunya Liu:** Writing - original draft, Conceptualization. **Yumei Wei:** Supervision, Formal analysis. **Qibin Zhang:** Visualization, Software. **Jingyan Ma:** Software. All the authors reviewed the manuscript.

**Data availability**

The datasets analysed during the current study are available in the IHME repository, https://vizhub.healthdata.org/gbd-results/

**Declaration of competing interest**

The authors declare no competing interests.

**Acknowledgements**

This work was supported by the Gansu Provincial Education Department's Graduate Student "Innovation Star" Project(2025CXZX-237; 2025CXZX-249), the Fundamental Research Funds for the Central Universities(31920240117; 31920250001; 31920230041), the Leading Talents Project of State Ethnic Affairs Commission of China and the Innovation Team of Ecosystem Restoration Modeling Theory and Application of Northwest Minzu University.

# Impact of the WHO's 90-70-90 Strategy on HPV-Related Cervical Cancer Control: A Mathematical Model Evaluation in China

—Supplemental Material—

Hua Liu, Chunya Liu, Yumei Wei, Qibin Zhang, Jingyan Ma

## Supplementary Note 1: Stability Diagram of the Disease-Free Equilibrium

We choose the following set of parameters to verify the theoretical analysis. $\Lambda = 1000$, $u = 0.01$, $\sigma = 0.01$, $d = 0.00001$, $\eta = 0.01$, $\omega = 0.05$, $\beta_1 = 0.2$, $\beta_2 = 0.1$, $\phi = 0.1$, $\varepsilon = 0.8$, $\tau = 0.001$, $\theta = 0.46674$, $\mu = 0.003$. Fig. S1 shows the stability of the DFE.

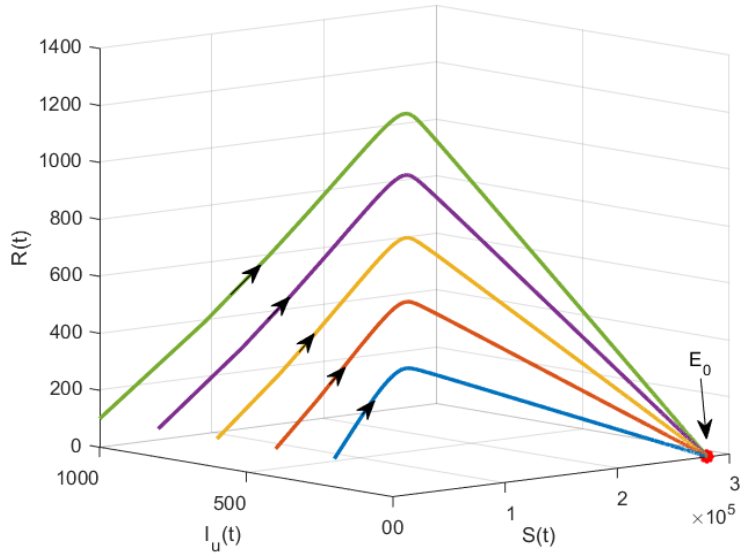

**Fig. S1** Stability of the DFE. The basic reproduction number $R_0 = 0.3683 < 1$.

## Supplementary Note 2: Endemic Equilibrium Point

To analyze the endemic equilibrium $E^* = (V^*, S^*, I_u^*, P^*, C^*, R^*)$, , we set the last three equations in Model (1) to zero,

$$\begin{cases} \theta\phi I_u^* - (\varepsilon + \tau + \mu)P^* = 0, \\ \theta\sigma I_u^* + \tau P^* - (d + \mu)C^* = 0, \\ \theta(1 - \phi - \sigma)I_u^* + \varepsilon P^* - \mu R^* = 0. \end{cases} \quad (1)$$

We have,

$$P^* = \frac{\theta\phi}{\varepsilon+\tau+\mu}I_u^*, \quad C^* = \frac{1}{d+\mu}(\theta\sigma + \frac{\tau\theta\phi}{\varepsilon+\tau+\mu})I_u^*,$$
$$R^* = \frac{1}{\mu}[\theta(1-\phi-\sigma) + \frac{\varepsilon\theta\phi}{\varepsilon+\tau+\mu}]I_u^*. \tag{2}$$

Add the second and third equations of model (1), we then have,

$$\begin{cases} (1-u)\Lambda + \omega V^* - (\eta+\mu)S^* - (\theta+\mu)I_u^* = 0, \\ u\Lambda + \eta S^* - (\omega+\mu)V^* = 0. \end{cases} \tag{3}$$

Solve Eq. (3) as follows,

$$V^* = \frac{\Lambda(\mu u+\eta) - \eta(\theta+\mu)I_u^*}{\mu(\mu+\eta+\omega)}, \quad S^* = \frac{\Lambda(\omega+\mu-\mu u) - (\theta+\mu)(\omega+\mu)I_u^*}{\mu(\mu+\eta+\omega)}. \tag{4}$$

Set the third equation of model (1) to zero, we have:

$$\frac{\beta_1 I_u^* + \beta_2 P^*}{N}S^* = (\theta+\mu)I_u^*, \tag{5}$$

Substituting Eq. (3) and (4) into Eq. (5), we can get,

$$\begin{aligned}
(\beta_1 + \beta_2 \frac{\theta\phi}{\varepsilon+\tau+\mu}) & \left[\frac{\Lambda(\omega+\mu-\mu u) - (\theta+\mu)(\omega+\mu)I_u^*}{\mu(\mu+\eta+\omega)}\right] \\
= (\theta+\mu) & \left[\frac{\Lambda(\mu u+\eta) - \eta(\theta+\mu)I_u^*}{\mu(\mu+\eta+\omega)} + \frac{\theta\phi}{\varepsilon+\tau+\mu}I_u^* \right. \\
& + \frac{\Lambda(\omega+\mu-\mu u) - (\theta+\mu)(\omega+\mu)I_u^*}{\mu(\mu+\eta+\omega)} + I_u^* \\
& \left. + \frac{1}{d+\mu}(\theta\sigma + \frac{\tau\theta\phi}{\varepsilon+\tau+\mu})I_u^* + \frac{1}{\mu}[\theta(1-\phi-\sigma) + \frac{\varepsilon\theta\phi}{\varepsilon+\tau+\mu}]I_u^*\right],
\end{aligned} \tag{6}$$

simplify Eq. (6) as follows,

$$\begin{aligned}
\frac{\Lambda}{\mu}R_0 - R_0\frac{(\theta+\mu)(\omega+\mu)I_u^*}{\mu(\omega+\mu-\mu u)} = & \frac{\Lambda}{\mu} - \frac{\theta+\mu}{\mu}I_u^* + I_u^* + \frac{\theta\phi}{\varepsilon+\tau+\mu}I_u^* + \frac{\tau\theta\phi}{(d+\mu)(\varepsilon+\tau+\mu)}I_u^* \\
& + \frac{\varepsilon\theta\phi}{\mu(\varepsilon+\tau+\mu)}I_u^* + \frac{\theta\sigma}{d+\mu}I_u^* + \frac{\theta(1-\phi-\sigma)}{\mu}I_u^*,
\end{aligned} \tag{7}$$

hence,

$$I_u^* = \frac{\Lambda(R_0 - 1)}{R_0\frac{(\theta+\mu)(\omega+\mu)}{\omega+\mu-\mu u} - \frac{d\theta[\phi\tau + \sigma(\varepsilon+\tau+\mu)]}{(d+\mu)(\varepsilon+\tau+\mu)}}. \tag{8}$$

## Supplementary Note 3: Stability Diagram of the Endemic Equilibrium

We choose the following set of parameters to verify the theoretical analysis. $\Lambda = 1000$, $u = 0.01$, $\sigma = 0.01$, $d = 0.00001$, $\eta = 0.01$, $\omega = 0.05$, $\beta_1 = 0.2$, $\beta_2 = 0.1$, $\phi = 0.1$, $\varepsilon = 0.8$, $\tau = 0.001$, $\theta = 0.01687$, $\mu = 0.003$. Fig. S2 shows the stability of the EE.

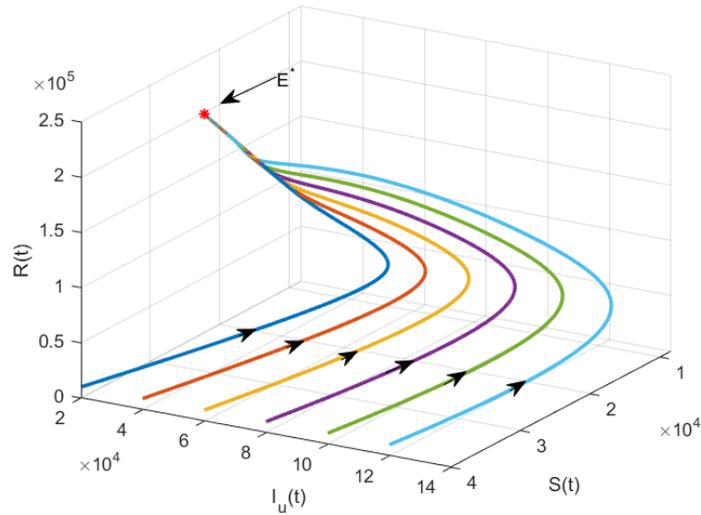

**Fig. S2** Stability of the EE. The basic reproduction number $R_0 = 8.4699 > 1$.

## Supplementary Note 4: Estimated Parameter Values for Model (1)

**Table S1** Model (1) parameters values and initial values with 95% confidence intervals

| Parameter | Mean value | Std | Source |
|---|---|---|---|
| $\Lambda$ | 3,215,701 | – | (ii) |
| $\mu$ | 1/79.43 | – | (iii) |
| $\eta$ | 0.00098 | 0.00051 | MCMC |
| $\omega$ | 0.04133 | 0.00475 | MCMC |
| $u$ | 0.15963 | 0.08463 | MCMC |
| $\beta_1$ | 0.40850 | 0.04096 | MCMC |
| $\beta_2$ | 0.10697 | 0.02230 | MCMC |
| $\sigma$ | 0.00634 | 0.00157 | MCMC |
| $\theta$ | 0.35502 | 0.02701 | MCMC |
| $\varphi$ | 0.14241 | 0.08653 | MCMC |
| $\varepsilon$ | 0.83167 | 0.07446 | MCMC |
| $\tau$ | 0.00574 | 0.00378 | MCMC |
| $d$ | 0.00035 | 0.00032 | MCMC |
| $V(0)$ | 0 | – | Estimation |
| $S(0)$ | – | – | MCMC |
| $I_u(0)$ | – | – | MCMC |
| $P(0)$ | – | – | MCMC |
| $C(0)$ | – | – | MCMC |
| $R(0)$ | – | – | MCMC |